\documentclass[prd,nofootinbib,floats,aps,superscriptaddress,onecolumn]{revtex4}
\pdfoutput=1 
\usepackage{graphicx}
\usepackage{amssymb}
\usepackage{amsmath}
\usepackage{amsfonts}
\usepackage[usenames]{color}
\usepackage{amsbsy}
\usepackage{epsfig}
\newcommand{\cc}{{\mbox{c.c.\,}}}

\newcommand{\cU}{\mathcal U}
\newcommand{\cV}{\mathcal V}

\newcommand{\bkk}{\mathbf k}

\def\tr{\,{\rm tr}\,}
\newcommand{\fract}[2]{{\textstyle\frac{#1}{#2}}}

\newcommand{\half}{\frac{1}{2}} \newcommand{\p}{\partial}

\newcommand{\be}{\begin{equation}}
\newcommand{\ee}{\end{equation}}
\newcommand{\bea}{\begin{eqnarray}}
\newcommand{\eea}{\end{eqnarray}}
\newcommand{\ben}{\begin{enumerate}}
\newcommand{\een}{\end{enumerate}}
\newcommand{\bit}{\begin{itemize}}
\newcommand{\eit}{\end{itemize}}
\newcommand{\eq}[1]{eq.~(\ref{#1})}

\newcommand{\Eq}[1]{Eq.~(\ref{#1})}
\numberwithin{equation}{section}

\newcommand\brak[1]{\ensuremath{\bigl| #1\bigr\rangle}}
\newcommand\krab[1]{\ensuremath{\bigl\langle #1\bigr|}}

\def\a{\alpha}
\def\b{\beta}

\def\k{\kappa}

\newcommand{\threej}[6]{\ensuremath{ \left( \begin{smallmatrix} 
{#1}&{#2}&{#3}\\{#4}&{#5}&{#6} \end{smallmatrix} \right) }}

\newcommand{\cE}{\ensuremath{{\cal E}}}
\definecolor{BrickRed}{cmyk}{0,0.89,0.94,0.28}
\definecolor{MidnightBlue}{cmyk}{0.98,0.13,0,0.43}
\definecolor{DarkGreen}{rgb}{0,0.7,0.1}

\newcommand{\bfx}{{\bf x}} 
\newcommand{\bfX}{{\bf X}} 

\newcommand{\bfz}{{\bf z}}
\newcommand{\bfk}{{\bf k}}

\newcommand{\cC}{{\mathcal C}}
\newcommand{\cF}{{\mathcal F}}
\newcommand{\cEC}{{\mathcal E}[{\mathcal C}]} 
\newcommand{\cS}{{\mathcal S}}
\newcommand{\cD}{{\mathcal D}}
\newcommand{\cT}{{\mathcal T}}
\newcommand{\cG}{{\mathcal G}}
\newcommand{\cO}{{\mathcal O}}

\newcommand{\phicl}{{\ensuremath{\phi_{\rm cl}}}}

\begin{document}

\preprint{MIT-CTP 3885}
\title{Casimir Forces between Compact Objects:  I.  The Scalar Case}
 
\author{T.\ Emig}

\affiliation{Institut f\"ur Theoretische Physik, Universit\"at zu
  K\"oln, Z\"ulpicher Strasse 77, 50937 K\"oln, Germany}
\affiliation{Laboratoire de Physique Th\'eorique et Mod\`eles
  Statistiques, CNRS UMR 8626, B\^at.~100, Universit\'e Paris-Sud, 91405
  Orsay cedex, France}

\author{N.\ Graham}
\affiliation{Department of Physics, Middlebury College,
Middlebury, VT  05753} 
\affiliation{Center for Theoretical Physics, Laboratory for Nuclear Science, and
Department of Physics, Massachusetts Institute of
Technology, Cambridge, MA 02139, USA}

\author{R.\ L.\ Jaffe}
\affiliation{Center for Theoretical Physics, Laboratory for Nuclear Science, and
Department of Physics, Massachusetts Institute of
Technology, Cambridge, MA 02139, USA}

\author{M.\ Kardar}
\affiliation{Department of Physics, Massachusetts Institute of
Technology, Cambridge, MA 02139, USA} 

\begin{abstract} 
  We have developed an exact, general method to compute Casimir
  interactions between a finite number of compact objects of arbitrary
  shape and separation. Here, we present details of the method for a
  scalar field to illustrate our approach in its most simple form; the
  generalization to electromagnetic fields is outlined in
  Ref.~\cite{Emig:2007cf}.  The interaction between the objects is
  attributed to quantum fluctuations of source distributions on their
  surfaces, which we decompose in terms of multipoles.  A functional
  integral over the effective action of multipoles gives the resulting
  interaction. Each object's shape and boundary conditions enter the
  effective action only through its scattering matrix.  Their
  relative positions enter through universal translation matrices that
  depend only on field type and spatial dimension. The distinction of
  our method from the pairwise summation of two-body potentials is
  elucidated in terms of the scattering processes between three
  objects.  To illustrate the power of the technique, we consider
  Robin boundary conditions $\phi -\lambda \partial_n \phi=0$, which
  interpolate between Dirichlet and Neumann cases as $\lambda$ is
  varied.  We obtain the interaction between two such spheres
  analytically in a large separation expansion, and numerically
  for all separations. The cases of unequal radii and unequal
  $\lambda$ are studied.  We find sign changes in the force as a
  function of separation in certain ranges of $\lambda$ and see
  deviations from the proximity force approximation even at short
  separations, most notably for Neumann boundary conditions.
\end{abstract}

\rightline{\small MIT-CTP-3885}

\maketitle

\section{Introduction}
\label{introduction}

Casimir forces arise when the quantum fluctuations of a scalar,
vector, or even fermion field are modified by the presence of static
or slowly changing external objects~\cite{Casimir:1948dh}.  The
objects can be modeled by boundary conditions that they place on {the
  fluctuating field $\phi$}, by an external field, $\sigma$, to which
$\phi$ couples~\cite{sigmawork}, or, in the case of electromagnetism,
by a material with space and frequency dependent dielectric and
magnetic properties.  The Casimir energy is the difference between the
energy of the fluctuating field {when then objects are present} and
when the objects are removed to infinite separation.

The advent of precision experimental measurements of Casimir forces
\cite{Bordag:2001qi} and the possibility that they can be applied to
nanoscale electromechanical devices~\cite{chan,Capasso:2007nq} has
stimulated interest in developing a practical way to calculate the
dependence of Casimir energies on the shapes of the objects.  Many
geometries have been analyzed over the years, but the case of compact
objects has proved rather difficult.  In a recent Letter
\cite{Emig:2007cf} we described a new method that makes possible
accurate and efficient calculations of Casimir forces and torques
between any number of compact objects.  The method applies to
electromagnetic fields and dielectrics as well as perfect conductors.
It {also} applies to other fields, {such as} scalar and Dirac, and to
any boundary conditions.  {In this approach,} the Casimir energy is
given in terms of the {fluctuating field's} scattering amplitudes from
the individual objects, {which encode} the effects of the shape and
boundary conditions.  The scattering amplitudes are known analytically
in some cases and numerically in others.  If the scattering amplitudes
are known, then the method can be applied from asymptotically large
separation down to separations that are a small fraction of the
dimension of the objects.  Results at large separations are obtained
using low frequency and low angular momentum {expansions of}
scattering amplitudes.  The coefficients multiplying the successive
orders in inverse separation can be identified with increasingly
detailed characteristics of the objects.  At small separations the
manipulation of large matrices, whose dimensions grow with angular
momentum, eventually slows down the calculation.  However at these
distances other methods, notably the ``proximity force approximation''
{(PFA)}, apply.  Thus it is now possible to obtain an understanding of
Casimir forces and torques at all separations for compact objects.

The aim of this paper is to provide a pedagogical introduction to our
methods by treating in detail the simplest case, a scalar field
obeying a boundary condition on a sharp surface.  The complications of
electromagnetism and smoothly varying dielectrics were already
introduced briefly in {Ref.}~\cite{Emig:2007cf}.  They will be treated
in more detail in subsequent publications \cite{forthcoming}.  Our
approach relies on a marriage of methods from path integral and
scattering formalisms, so we provide background on both of these
subjects as they apply to Casimir effects.

Schwinger, in particular, emphasized that Casimir forces could be
understood as ordinary electromagnetic interactions between quantum
fluctuations of charge and current in metals or
dielectrics~\cite{schwinger}.  We implement this idea here through a
functional integral formulation.  We begin by writing the Casimir
energy as the logarithm of a functional integral over all field
fluctuations constrained by the boundary conditions on a set of
surfaces.  Following Refs.~\cite{Bordag+85,LK91} we implement the
boundary conditions by introducing integrals over sources that enforce
the boundary conditions as functional $\delta$-functions.  Next we
perform the functional integral over the field.  The result is an
integral of the form $\prod_{\a}\int {\cD}\varrho_\alpha
\exp(iS[\varrho])${,} where $\varrho_\alpha$ are the sources on the
different surfaces labeled by $\alpha$ and $S[\varrho]$ is the
classical action of the field{, which is} uniquely determined by the
source{s} $\varrho_\alpha$.  The contributions to $S[\varrho]$ are of
two qualitatively different forms, those involving sources on
different surfaces and those coupling a particular surface source to
itself through the field it generates.  Both have simple
representations in a basis of angular momentum eigenstates (``partial
waves'') and multipole moments of the sources.  The couplings between
different objects involve only a well-known kinematic ``translation
matrix,'' $\mathbb{U}$, that relates a partial wave amplitude
generated by one source to partial waves seen by
another~\cite{wittmann}.  The self-interactions depend only on the
``transition matrix,'' $\mathbb{T}$, (related to the scattering
matrix, $\mathbb{S}$, by
$\mathbb{T}=\half(\mathbb{S}-\mathbb{I})$])~\cite{newton}, which
describes the response of the scalar field to the boundary condition
on the surface.  The result is remarkably simple.  For a complex
scalar field in the presence of two objects it takes the form,
\begin{equation}
\label{tform}
\cE_{12}[\cC]=\frac{\hbar c}{\pi}\int_0^\infty d\kappa \,
\ln\det(\mathbb{I}-\mathbb{T}^{1}\mathbb{U}^{12}
\mathbb{T}^{2}\mathbb{U}^{21}) \, ,
\end{equation}
where the determinant is over the partial wave indices on the matrices
$\mathbb{T}^\alpha$ and $\mathbb{U}^{\alpha\beta}$ and the integral is over
$\kappa=-i\omega/c$, the imaginary wavenumber.  In Section \ref{appl}
the usefulness of this result is demonstrated through several specific
applications.

Casimir forces between compact objects were first considered by
Casimir and Polder in 1948~\cite{casimirpolder}.  Since then, two
threads of work related to ours have been pursued: attempts to
evaluate the Casimir force between compact objects explicitly, and
efforts to develop a general framework similar to that embodied in
Eq.~\eqref{tform}.  Until recently work along the first line consisted
of expansions at asymptotically large separation.  Casimir and Polder
found the electromagnetic force between two neutral but polarizable
atoms to leading order at large separation~\cite{casimirpolder}.
Feinberg and Sucher~\cite{fs} generalized to arbitrary compact objects
and included magnetic effects.  Balian and Duplantier studied perfect
metals, and derived explicit results to leading order at
asymptotically large separation~\cite{Balian:1977qr}.  More recently
Gies et al.~\cite{Gies:2003cv} used numerical methods to evaluate the
Casimir force between two Dirichlet spheres for a scalar field, over a
range of subasymptotic separations{, and in other open geometries
  {such} as a plate and a cylinder \cite{Gies+06b} or finite plates
  with edges \cite{Gies+06a}.}  Bulgac and
collaborators~\cite{Bulgac:2005ku} applied scattering theory methods
to the same {scalar Dirichlet} problem and obtained results over a
wide range of separations.  The only explicit calculations for
subasymptotic distances up to now have been for a scalar field obeying
Dirichlet boundary conditions on two spheres, a sphere and a
plate~\cite{Bulgac:2005ku} and for electromagnetic fields for a
  plate and a cylinder \cite{Emig+06a} and two perfectly conducting
  spheres \cite{Emig:2007cf}.

Formulas for the Casimir energy closely related to Eq.~\eqref{tform}
have appeared previously in the literature.  The first appearance we
are aware of is in the work of Balian and Duplantier
\cite{Balian:1977qr} based on the multiple reflection expansion (MRE).
Their Eq.~(7.20) gives a quantity, $\Psi$, that is directly related to
the Casimir energy, and could form the basis for an approach like
ours.  Their approach, like ours, punctuates periods of propagation in
the vicinity of {each} object with free propagation between the
objects. However no explicit expressions for the propagation kernels,
$K_{\a}$, are given.  More recently ``log-det'' formulas like ours
have arisen in several functional integral studies of Casimir
forces~\cite{Emig+04,Emig+06a}.  To our knowledge Kenneth and Klich in
Ref.~\cite{Kenneth:2006vr} were the first to identify the inverted
Green's function as a $\mathbb{T}$-matrix and derive Eq.~\eqref{tform}
as a formal result.  Here also, no explicit form for the matrices was
derived.  Their derivation applies to scalar fields in a medium with a
space and frequency dependent speed of light.  For the case of a
scalar field, our result can be viewed as an explicit {expression for}
their formula in a basis of partial waves, which we show is
particularly well suited to practical applications.  Our new
derivation of Eq.~\eqref{tform} in terms of quantum fluctuations of
sources has the advantage that it allows for a physically transparent
extension to gauge fields in the presence of material objects with
general dielectric and magnetic properties, as outlined in
Ref.~\cite{Emig:2007cf}.

In Section \ref{appl}, we provide a number of examples where our
approach yields straightforward results for Casimir interactions that
could not be obtained before. We {carry out} explicit analytical and
numerical computations for spheres with Robin boundary conditions
$\phi-\lambda\partial_n\phi=0$. The sign of the Casimir interaction at
asymptotically large and small distances is classified {as a
  function of} the value of the Robin parameter $\lambda$.  The
Casimir energy at asymptotically large separations is obtained as a
series in the ratio of distance to sphere radius.  We also perform
numerical computations of the Casimir interaction of two spheres with
Robin boundary conditions spanning the full range of separations,
including very short distances, where our results confirm the
proximity force approximation (PFA).  At intermediate distances
considerable deviations from the PFA are found. For certain values of
$\lambda$, the Casimir force can change sign (once or twice) as
function of separation.

Finally, in Sec.~\ref{sec:inter-terms-low}, we give an expression for
the Casimir interaction at large separations between two compact
object of arbitrary shape and with  arbitrary boundary conditions in
terms of generalized capacitance coefficients.  {Technical
derivations are presented in three Appendices.}

\section{Foundations}
\label{foundations}

In this Section we review  formalism essential for our work.  We
use both functional integral methods and techniques originating in
scattering theory.  First, in Sec.~\ref{2.1} we introduce the
functional integral approach of Refs.~\cite{LK91,Emig+01,Buscher+05},
which allows us to trade the problem of fields fluctuating in the bulk
for the interactions of sources defined only on the bounding surfaces.
In our approach, individual objects are characterized by the way they
scatter the {fluctuating} fields.  This information is summarized in the
transition matrix, $\mathbb{T}$, which we introduce in
Sec.~\ref{2.2} and relate to scattering solutions of the equations of
motion and to Green's functions.  In the $k\to 0$ limit,
Helmholtz's equation reduces to Laplace's equation.  In this limit the
$\mathbb{T}$-matrix can be related to generalized coefficients of
capacitance, which we summarize in Sec.~\ref{2.3}.
Finally in Sec.~\ref{2.4} we introduce  the
translation formulas that relate partial waves computed with respect
to one origin to those computed with respect to another.

\subsection{Functional integral formulation}
\label{2.1}
We consider a complex quantum field, $\phi(\bfx,t)$, which is defined
over all space {and} constrained by boundary conditions
{$\cal C$} on a set of fixed surfaces $\Sigma_\alpha$, for
$\alpha=1,2, \dots, N$, but is otherwise non-interacting.  We assume that the
surfaces are closed and compact  and refer to their interiors as
``objects.''  Our starting point is the functional integral
representation for the trace of the propagator, ${\rm Tr}\,
e^{-iH_{\cC}T/\hbar}$~\cite{fandh},
\begin{equation}
\label{effact}
{{\rm Tr}\, e^{-iH_{\cC}T/\hbar}=\int \left[{\cal D}\phi\right]_{\cal C}  
\, e^{\frac{i}{\hbar}S_{E}[\phi]}\,\equiv\, Z\,[\cC] \, ,}
\end{equation}
where the subscript $\cC$ denotes the constraints imposed by the
boundary conditions.\footnote{We have used an abbreviated
  notation for the functional integral.  Since $\phi$ is complex
  $\int\cD\phi$ should be understood as $\int\cD\phi\cD\phi^{*}$, and
  similarly in subsequent functional integrals.}
The integral is over all field {configurations} that obey
the boundary conditions and are periodic in a time interval $T$.
$S[\phi]$ is the action for a free complex field,
\begin{equation}
\label{action1}
S_{E}[\phi]= \int_{0}^{T}dt\int d\bfx\,
\left({\frac{1}{c^2}}|\partial_t\phi|^{2}-|{\nabla}\phi|^{2}\right) ,
\end{equation}
where the $\bfx$-integration covers all space.\footnote{Note that
  $\phi$ is defined and can fluctuate inside the objects bounded by
  the surfaces $\Sigma_\alpha$.  In this feature our formalism departs
  from some treatments where the field is defined to be strictly zero
  (for Dirichlet boundary conditions) inside the objects.  The
  fluctuations interior to the objects do not depend on the
  separations between them and therefore do not affect Casimir forces
  or torques.}
 
The ground state energy can be projected out of the trace in 
Eq.~\eqref{effact} by setting $T=-i\Lambda/c$ taking the limit
$\Lambda\to\infty$,
\begin{equation}
\label{proj}
{\cE}_{0}[\cC] = -\lim_{\Lambda\to\infty}\frac{\hbar c}{\Lambda}
\ln\left({\rm Tr \,} 
e^{-H_{\cC}\Lambda/\hbar c}\right)=-\lim_{T\to\infty}
\frac{\hbar c}{\Lambda} \ln Z[\cC] \, ,
\end{equation}
and the Casimir energy is obtained by subtracting the ground state
energy when the objects have been removed to infinite separation,
\begin{equation}
\label{casimir1}
\cEC = -\lim_{\Lambda\to\infty}\frac{\hbar c}{\Lambda}
\ln\left( Z[\cC]/ Z_{\infty}\right) \, .
\end{equation}
In the standard formulation, the constraints are implemented by
boundary conditions on the field $\phi$
at the surfaces $\{\Sigma_\alpha\}$.  The usual
choices are Dirichlet, $\phi=0$, Neumann, $\p_{n}\phi=0$, or mixed
{(Robin)}, $\phi {-} \lambda\p_{n}\phi=0$, where  $\p_{n}$
is the normal derivative pointing out of the objects.  To be
specific, we first consider Dirichlet boundary conditions.
The extension to the Neumann case is presented in Appendix
\ref{app:B}.  As noted in the Introduction, the only effect of the
choice of boundary conditions is to determine which $\mathbb{T}$-matrix
appears in the functional determinant, Eq.~\eqref{tform}.

For readers who are not familiar with the functional integral
representation of the Casimir energy, Eq.~\eqref{casimir1}, in
Appendix \ref{app:A} we show that Eq.~\eqref{casimir1} agrees with the
traditional definition of the Casimir energy in terms of zero-point
energies of normal modes.

Since the constraints on $\phi$ are time independent, the integral
over $\phi(\bfx, t)$ may be written as an infinite product of
integrals over Fourier components,
\begin{equation}
\int \left[{\cal D}\phi\right]_{\cal C} =
\prod_{n=-\infty}^{\infty}\left[{\cD}\phi_{n}(\bfx)\right]_{\cC} \, ,
\end{equation}
where 
\begin{equation}
\phi(\bfx, t)=\sum_{n=-\infty}^{\infty}\phi_{n}(\bfx)e^{2\pi i n t/T} ,
\end{equation}
and the logarithm  of $Z$ becomes a sum, 
\begin{equation}
\label{firstlog}
\ln Z[\cC] = \sum_{n=-\infty}^{\infty}
\ln\left\{\int\left[{\cD}\phi_{n}(\bfx)\right]_{\cC} 
\exp \left[ i\frac{T}{\hbar}\int d\bfx
\left(\left(\frac{2\pi n}{{c}T}\right)^{2}|
\phi_{n}(\bfx)|^{2}- |\nabla\phi_{n}(\bfx)|^{2}  \right)\right]\right\} \, .
\end{equation}

As $T\to\infty$,  $\sum_n$ can be replaced by
$\fract{cT}{2\pi}\int_{-\infty}^{\infty}dk$, 
where $k=2\pi n /(cT)$ and $\phi_{n}(\bfx)$
is replaced by $\phi(\bfx,k)$.  Combining the positive and negative
$k$-integrals gives
\begin{eqnarray}
\label{secondlog}
\ln Z[\cC] &=&  \frac{cT}{\pi}\int_{0}^{\infty}d  k\ln 
\left\{\int\left[{\cD}\phi (\bfx,  k)\right]_{\cC} 
\exp \left[ i\frac{T}{\hbar}\int d\bfx\left(  k^{2}|\phi(\bfx,  k)|^{2} -
|\nabla\phi(\bfx,  k)|^{2}\right)\right]\right\}
\nonumber\\
&=&\frac{cT}{\pi}\int_{0}^{\infty}d  k\ln\, \mathfrak{Z}_{\cC}(  k) \, ,
\end{eqnarray}
where
\begin{equation}
\label{minkowski-z}
\mathfrak{Z}_{\cC}( k) =\int\left[{\cD}\phi (\bfx,k)\right]_{\cC} 
\exp \left[ i\frac{T}{\hbar} \int d\bfx\left(k^{2}|\phi(\bfx,k)|^{2} -
|\nabla\phi(\bfx,k)|^{2}\right)\right],
\end{equation}
is the functional integral at fixed $k$.

To extract the Casimir energy, we use $T=-i\Lambda/c$ and Wick rotate
the $k$-integration ($k=i\kappa$ with $\kappa>0$){.}
\footnote{A more
careful treatment of the rotation of the integration contour to the
imaginary axis is necessary in the presence of bound states.}  Using
Eq.~\eqref{casimir1}, we obtain,
\begin{equation}
\label{casimir2}
\cEC = {-} \frac{\hbar c}{\pi}\int_{0}^{\infty} d\kappa
\ln\frac{\mathfrak{Z}_{C}(i\kappa)}{\mathfrak{Z}_{\infty}(i\kappa)} \, .
\end{equation}
{Here} $\mathfrak{Z}_{\cC}(i\k)$ is given by the Euclidean
functional integral,
\be
\label{euclid-z}
\mathfrak{Z}_{\cC}(i\kappa) =\int\left[{\cD}\phi (\bfx,i\k)\right]_{\cC} 
\exp \left[ - \frac{T}{\hbar} \int d\bfx\left(\k^{2}|\phi(\bfx,i\k)|^{2}
+ |\nabla\phi(\bfx,i\k)|^{2}\right)\right]\, .
\ee

It remains to incorporate the constraints directly into the functional
integral using the methods of Refs.~\cite{Bordag+85,LK91}.  Working in Minkowski
space, we consider the fixed frequency functional
integral, $\mathfrak{Z}_{\cC}(k)$ (and suppress the label $k$ on the
field $\phi$).  Following Ref.~\cite{Bordag+85,LK91}, we implement the
constraints in the functional integral by means of a functional
$\delta$-function.   For Dirichlet boundary conditions
the constraint reads,
\begin{equation}
\label{deltafn}
\int \left[{\cD}\phi (\bfx)\right]_{\cC}=\int \left[{\cD}\phi (\bfx)\right] 
\prod_{\a=1}^{N}\int \left[{\cD}\varrho_{\a}(\bfx)\right]
\exp\left[ i \frac{T}{\hbar}\int_{\Sigma_{\a}} 
d\bfx \left(\varrho^{*}_{\a}(\bfx)\phi(\bfx)+ \cc\right)\right] \, ,
\end{equation}
where the functional integration over  $\phi$ is no longer
constrained.  Other boundary conditions can be
implemented similarly.  In the resulting functional integral,
\begin{eqnarray}
\mathfrak{Z}_{\cC}(k)&=&\prod_{\a=1}^{N}\int
\left[{\cD}\varrho_{\a}(\bfx )
\right]\int \left[{\cD}\phi 
(\bfx)\right] \exp \left[ i\frac{T}{\hbar}\left( 
\int d\bfx\left(k^{2}|\phi(\bfx)|^{2}-|
\nabla\phi(\bfx)|^{2}  \right)+ \sum_{\a}\int_{\Sigma_{\a}} d\bfx
\left(\varrho^{*}_{\a}(\bfx )
\phi(\bfx )+ \cc\right)\right)\right]\nonumber\\
&\equiv&\prod_{\a=1}^{N}\int \left[{\cD}\varrho_{\a}(\bfx )\right]\int 
\left[{\cD}\phi (\bfx)\right] 
\exp \left(i\frac{T}{\hbar}\widetilde S[\phi,\varrho]\right) \, ,
\label{source3}
\end{eqnarray} 
the fields fluctuate without constraint throughout space and the
sources $\{\varrho_{\a}\}$ fluctuate on the surfaces.  We denote the new
``effective action'' including both the fields and sources by
$\widetilde S[\phi,\varrho]$.

\subsection{The scattering amplitude}
\label{2.2}

In this section we consider each of the objects  in
isolation, and review the solution to the Helmholtz equation
\begin{equation}
\label{helmh}
-\left(\nabla^{2}+k^{2}\right)\phi_{\alpha}(\bfx)=0,
\end{equation}
in the domain \emph{outside} $\Sigma_{\alpha}$.  We assume that
$\phi_{\a}$ obeys the Dirichlet boundary condition on $\Sigma_{\a}$.
It is convenient to fix an origin, $\cO_{\a}$, at some point inside
the object and introduce a coordinate vector, $\bfx_{\a}$, defined
with respect to this origin.  A specific choice of the origin will be
made later to simplify the analysis.  For simplicity we suppress the
label $\a$ from now on in this section.  We introduce spherical polar
coordinates relative to this origin, $r\equiv |\bfx|$ and $\hat\bfx$.
Outside $\Sigma$, $\phi$ is a superposition of partial waves with
definite angular momentum,
\begin{equation}
\phi(\bfx)=\sum_{lm}\left[c_{lm}h_{lm}^{(2)}(kr)+d_{lm}h_{lm}^{(1)}(kr)\right]
Y_{lm}(\hat\bfx) \, ,
\end{equation}
where the spherical Hankel functions, $h_{lm}^{(1)}(kr)$ and
$h_{lm}^{(2)}(kr)$ are the outgoing (asymptotic to $e^{ikr}{/kr}$) and
incoming (asymptotic to $e^{-ikr}{/kr}$) solutions respectively.  To
make contact with traditional methods of scattering theory, we rewrite
$\phi(\bfx)$ in terms of solutions with unit incoming amplitude,
\begin{equation}
\label{general}
\phi(\bfx) =\sum_{lm}c_{lm}\phi_{lm}(\bfx) ,
\end{equation}
where 
\begin{equation}
\label{reg}
\phi_{lm}(\bfx) = h_{l}^{(2)}(k r)Y_{lm}(\hat\bfx)
+\sum_{l'm'}{\cal S}_{l'm'l m }(k )h_{l'}^{(1)}(k r)Y_{l'm'}(\hat\bfx) \, .
\end{equation}
The coefficients $\cS_{lml'm'}(k)$ measure the response of the object,
$\Sigma$, to a unit amplitude incoming wave with angular momentum
$(lm)$.  They are the matrix elements of the scattering operator, or
$\mathbb{S}$-matrix,
\begin{equation}
\cS_{l'm'lm}(k)=\krab{l'm'}\mathbb{S}(k)\brak{lm} \,,
\end{equation}
which are fixed by the condition that the
resulting basis functions, {$\phi_{lm}$} 
vanish on $\Sigma$.  With the object absent, $\mathbb{S}$
would go to the identity operator, $\cS_{l'm'lm}\to\delta_{ll'}\delta_{mm'}$,
and $\phi_{lm}(\bfx)$ would reduce to $2j_{l}(k r)Y_{lm}(\hat\bfx)$,
the partial wave solution regular at the origin.  It is convenient
to make this result explicit by rewriting Eq.~\eqref{reg} as
\begin{equation}
\label{prereg}
 {\phi_{lm}(\bfx)} = 2 j_{l}(k r)Y_{lm}(\hat\bfx)
+\sum_{l'm'}2\,{\cal T}_{l'm'l m }(k )h_{l'}^{(1)}(k r)Y_{l'm'}(\hat\bfx) \, ,
\end{equation}
where $\cT$ is the \emph{transition matrix} or
$\mathbb{T}$-matrix\footnote{Our definition of the $\mathbb{T}$-matrix
  differs by a factor $i$ from some conventional choices~\cite{schiff}.}
\begin{equation}
\cT_{l'm'lm}(k)\equiv \krab{l'm'}\mathbb{T}(k)\brak{lm}=\frac{1}{2}\,
\left(\cS_{l'm'lm}(k)-\delta_{l'l}\delta_{m'm}\right) \, .
\end{equation}
Unitarity requires $\mathbb{S}^{\dagger}\mathbb{S}=\mathbb{I}$, or
$\sum_{l'm'}\cS^{*}_{l'm'l''m''}\cS_{l'm'lm}
=\delta_{l''l}\delta_{m''m}$.  Time reversal {symmetry} requires
that $\cS_{l'm'lm}$ is symmetric in $(l,m)\leftrightarrow
(l',m')$.  If the object $\Sigma$ is spherically symmetric, $\mathbb{S}$ and
$\mathbb{T}$ are diagonal in $\ell$ and $m$.  However we are also
interested in  more general cases where the material objects do not
have any special symmetry.

It is useful to review the connection between $\cT_{l'm'lm}(k)$ and
the scattering of $\phi$ from the object.  The scattering amplitude is
defined as the response of the object to an incoming plane wave.
Since angular momentum is not conserved{,} we have to keep track of the
direction of the incoming plane wave.  The solution to the Helmholtz
equation that reduces at $r\to\infty$
to a plane wave with wave vector $\bfk$
accompanied by an outgoing scattered wave defines the
scattering amplitude, $f(\bfk,\bfk')$,
\begin{eqnarray}
\Phi(\bfk, \bfx)
&=& 2\pi \sum_{lm}\phi_{lm}(\bfx)i^{l}Y_{lm}^{*}(\hat\bfk)\nonumber\\
&\sim& e^{i\bfk\cdot \bfx} + 
\frac{e^{ik r}}{r}f(\bfk,\bfk')\quad\mbox{as }\, r\to \infty \, ,
\label{boundcond}
\end{eqnarray}
where $\bfk'\equiv k\,\hat\bfx$ is the wave vector of the observed
wave.  The differential cross section is given by
\begin{equation}
\label{crosssection}
\frac{d\sigma}{d\Omega}=\left|f(\bfk,\bfk')\right|^{2} \, .
\end{equation}
To relate $f(\bfk,\bfk')$ to $\mathbb{T}$ we express the plane wave a
a superposition of partial waves,
\begin{equation}
\label{pw}
e^{i\bfk\cdot\bfx} =4\pi \sum_{lm}j_{l}(k r)i^{l}
Y_{lm}(\hat\bfx) Y_{lm}^{*}(\hat\bfk) \, , 
\end{equation}
where  {the unit vectors} $\hat\bfx $ and $\hat\bfk$ {specify}
the polar angles of $\bfx$ and $\bfk$ with respect to an
arbitrary axis.  We identify what remains with the scattered wave,
\begin{eqnarray}
\label{scattsol}
\Phi(\bfk,\bfx) 
&=& e^{i\bfk\cdot \bfx}+4\pi \sum_{l'm'lm}i^{l}{\cal
T}_{l'm'lm}(k)h_{l'}^{(1)}(k r)Y_{l'm'}(\hat\bfx)
Y_{lm}^{*}(\hat\bfk) \, .
\end{eqnarray}
Using the asymptotic form of the Hankel functions, $h_{l}^{(1)}(k
r)\to i^{-l-1}e^{ikr}/kr$, we obtain
\begin{equation}
\label{scattamp}
f(\bfk,\bfk')=
\frac{4\pi}{ik}\sum_{l'm'lm}i^{l-l'}\cT_{l'm'lm}(k)
Y_{l'm'}(\hat\bfk')Y_{lm}^{*}(\hat\bfk) \, . 
\end{equation}
When angular momentum is conserved,
$\cS_{l'm'lm}(k) = e^{2i\delta_{l}(k)}{\delta_{ll'}\delta_{mm'}}$ and 
we obtain the standard result
$f(\bfk,\bfk')=\sum_{l}(2l+1)f_{l}(k
)P_{l}(\cos\theta_{\hat\bfk\hat\bfk'})$, with ${f_l(k)}
  =\frac{1}{k}\sin\delta_{l}(k )e^{i\delta_{l}(k)}$.

Finally, we will need the partial wave expansion for the \emph{free},
outgoing wave Helmholtz Green{'s} function,
\begin{equation}
\label{gf3}
{\cal G}_{0}(\bfx,\bfx',k) \equiv \frac{e^{ik|\bfx-\bfx'|}}{4\pi|\bfx-\bfx'|}
= {ik}\sum_{lm}j_{l}(k r_{<})
h^{(1)}_{l}(k r_{>})Y_{lm}(\hat\bfx)Y_{lm}^{*}(\hat\bfx')=
{ik}\sum_{lm}j_{l}(k r_{<})
h^{(1)}_{l}(k r_{>})Y_{lm}(\hat\bfx')Y_{lm}^{*}(\hat\bfx) \, ,
\end{equation}
where the notations $r_{<(>)}$ refer to whichever of $r,r'$ is the
smaller (larger).  

\subsection{Low energy scattering and the  coefficients of capacitance}
\label{2.3}

As $k\to 0$, the Helmholtz equation reduces to Laplace's equation.
Therefore we can relate the low energy limit of the
$\mathbb{S}$-matrix elements to the parameters that describe solutions
to Laplace's equation, which are tensor generalizations of
capacitance.  In this subsection we continue to suppress the label
$\alpha$ that distinguishes the particular object of interest.

At large distances, solutions to Laplace's equation are of the form
$r^{l}Y_{lm}(\hat\bfx)$ or $r^{-l-1}Y_{lm}(\hat\bfx)$.  If an object
is placed in an external field {(potential)} of the first form, {then}
its response, determined by the boundary condition $\phi=0$ on
the surface, is of the second form.  The monopole response to an
asymptotically constant field defines the capacitance; the dipole
response to an asymptotically dipole field defines the polarizability,
and so forth. Therefore we parameterize a solution to Laplace's
equation that goes to $r^{l}Y_{lm}(\hat\bfx)$ by\footnote{Note the
minus sign that preserves the usual definition of capacitance.  When
an object is held at a voltage $V$, its charge is $Q=CV$.
Equivalently, when the object is grounded and the potential at
infinity is $V$, the induced monopole field is $-CV/r$.}
\begin{equation}
\label{multipolesofphi}
\phi_{lm}(\bfx)\propto r^{l}Y_{lm}(\hat\bfx)-\sum_{l'm'}\cC_{l'm'lm}
\frac{Y_{l'm'}(\hat\bfx)}{r^{l'+1}} \, .
\end{equation}
For a Dirichlet boundary condition the $\{C_{l'm'lm}\}$ are the tensor
generalizations of the capacitance.  For other boundary conditions,
the physical connection to electrostatics is lost, but the tensor
structure (and the connection to the $\cS$-matrix) remains.

Taking the limit $k\to 0$ in Eq.~\eqref{prereg} we obtain an
expression for $\phi_{lm}$ in terms of the $k\to 0$ limit of the
$T$-matrix,
\begin{equation}
\label{ktozero}
\lim_{k\to 0}{\phi_{lm}(\bfx)} \sim \frac{2 (kr)^{l}}{(2l+1)!!}
Y_{lm}(\hat\bfx)
-{2i}\sum_{l'm'}{\cal T}_{l'm'lm}(k)\frac{(2l'-1)!!}{(kr)^{(l'+1)}}
Y_{l'm'}(\hat\bfx) \, .
\end{equation}
In order to obtain a finite and non-vanishing limit, we find
$\cT_{l'm'lm}(k)\sim k^{l+l'+1}$ as $k\to 0$, a standard result.
Comparing these two expressions we find
\begin{equation}
\label{eq:scatter}
\lim_{k\to 0}\frac{1}{k^{l+l'+1}}{\cal T}_{l'm'lm}(k)= 
\frac{-iC_{l'm'lm}}{(2l+1)!!(2l'-1)!!} \, .
\end{equation}

Some special cases deserve mention.  For $l=l'=0$, as $k\to 0$,
$\cT_{0000}(k)\sim-ika$, where $a$ is the scattering length.
Comparing Eqs.~\eqref{ktozero} and (\ref{eq:scatter}) we see that the
capacitance, $C\equiv C_{0000}$, equals the scattering length.  The
$l=0,l'=1$ coefficient describes the dipole moment developed by a
conductor in response to a constant external potential (or, by
symmetry, the charge induced by an external dipole field).  It
vanishes for spherically symmetric objects, but more generally we can
choose the origin of the coordinate system so that
$C_{1m00}=C_{00{1}m}=0$.  From now on we assume that the origin,
$\cO_{\alpha}$, within the object $\Sigma_{\alpha}$ has been chosen to
eliminate this dipole response.  The $l=l'=1$ coefficients describe
the dipole response to an external dipole field: these nine
components form the tensor (electrostatic) polarizability.

Finally, we note that the coefficients $C_{l'm'lm}$ can be interpreted
as the components of an abstract tensor $\mathbb{C}_{l'l}$ in a
spherical basis.  $\mathbb{C}_{l'l}$ is the exterior product of two
irreducible tensors, one rank $l$ and the other rank $l'$.  It is
sometimes useful to project $\mathbb{C}_{l'l}$ into its irreducible
components.  Thus, for example, $\mathbb{C}_{11}$, the electrostatic
polarizability, can be decomposed into its trace, the scalar
polarizability, and a traceless part that describes the aspherical
response of the object.

\subsection{Translation formulas}
\label{2.4}

\begin{figure}
\includegraphics[scale=0.25]{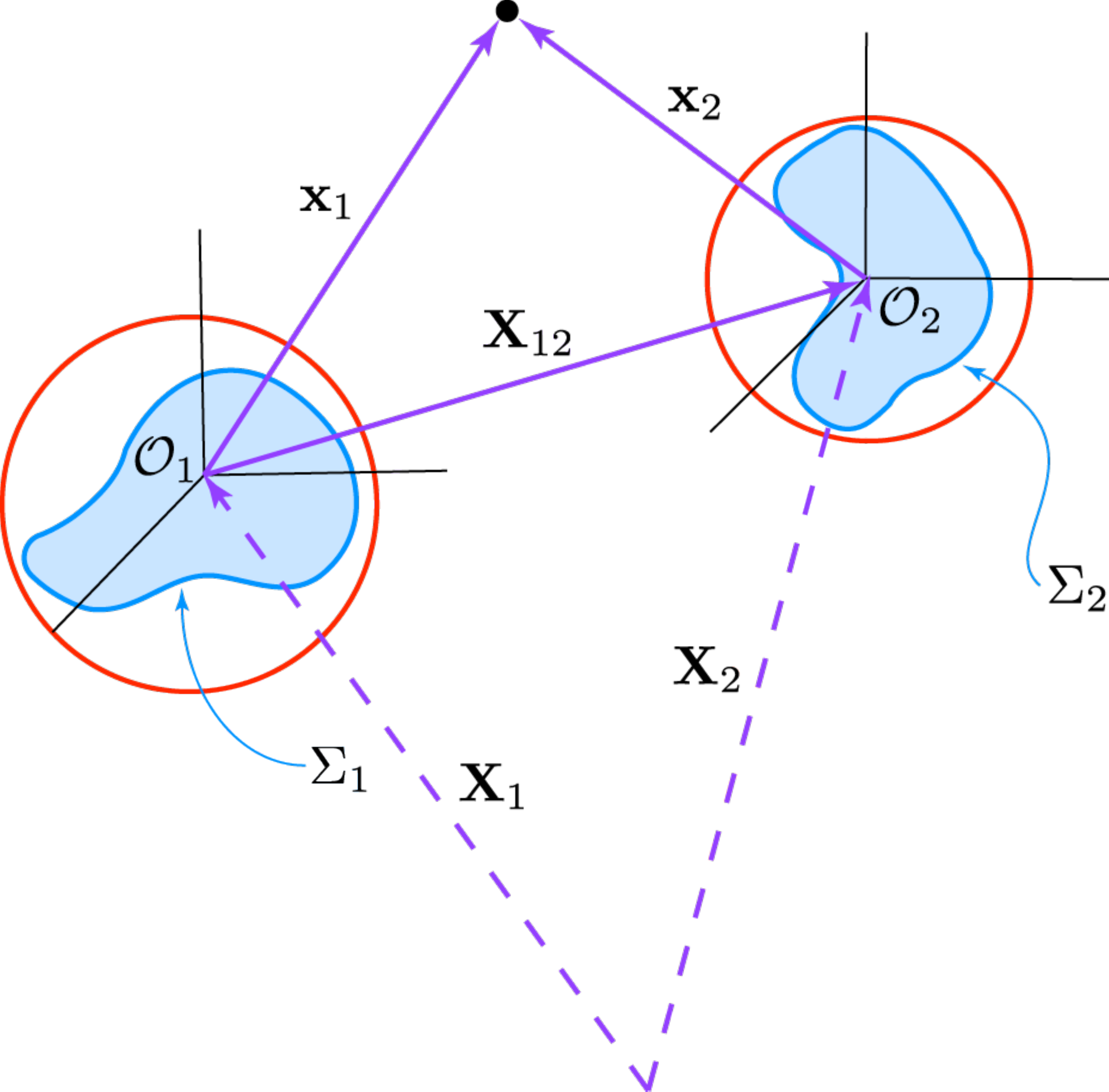}
\caption{(Color online) Two objects enclosed by the surfaces $\Sigma_{1}$ and
  $\Sigma_{2}$ are shown, each with bounding spheres (radii $R_{1}$
  and $R_{2}$).  We assume that it is possible to choose bounding
  spheres that do not overlap.  Coordinate systems with parallel axes
  are erected at origins $\cO_{1}$ and $\cO_{2}$ at positions $\bfX_1$
  and $\bfX_2$. The origins are chosen so that the dipole coefficients
  of capacitance, $C_{001m}$, vanish.  Coordinate vectors $\bfx_{1}$
  and $\bfx_{2}$ to an arbitrary point, $\bfx$, are shown.  The vector
  from $\cO_{1}$ to $\cO_{2}$ is
  $\bfX_{12}=\bfX_{2}-\bfX_{1}=\bfx_{1}-\bfx_{2}$.  The distance
  between the two objects is $d_{12}=|\bfX_{12}|$.}
\label{objects}
\end{figure}

Partial wave solutions to the Helmholtz equation for an object $\Sigma_{\a}$,
defined with respect to the origin $\cO_{\a}$, can be expanded in partial
waves with respect to a second origin $\cO_{\b}$ within a second
object $\Sigma_{\b}$.
  The objects, coordinate origins, and
notation for vectors  are shown in
Fig.~\ref{objects}.
Any solution with definite angular momentum defined with respect to
$\cO_{\a}$, whether regular ($j_{l}$) or outgoing ($h^{(1)}_{l}$), is
regular when viewed from $\cO_{\b}$.  Therefore it is possible to
expand both in terms of spherical Bessel functions
$j_{l}(kr_{\b})$ with respect to $\cO_{\b}$,
\begin{eqnarray}
  \label{eq:reg+out+solutions}
j_{l}(kr_{\a})Y_{lm}(\hat\bfx_{\a}) &=&
\sum_{l'm'}\cV_{l'm'lm}^{\b\a}(\bfX_{\b\a})
j_{l'}(kr_{\b})Y_{l'm'}(\hat\bfx_{\b})\nonumber\\
h^{(1)}_{l}(kr_{\a})Y_{lm}(\hat\bfx_{\a}) &=&
\sum_{l'm'}\cU_{l'm'lm}^{\b\a}(\bfX_{\b\a})
j_{l'}(kr_{\b})Y_{l'm'}(\hat\bfx_{\b}) 
\end{eqnarray}
for $r_{\b}<d_{\a\b}$.  Note that $\bfx_{\a}$ and $\bfx_{\b}$ refer to
the coordinate components of a point $\bfx$ relative to the origin
$\cO_{\a}$ or $\cO_{\b}$ respectively.  The matrices $\mathbb{U}$ and
$\mathbb{V}$, that define the change of basis depend on the vector
from $\cO_{\a}$ to $\cO_{\b}$, ${\bfX_{\a\b}}=\bfX_{\b}-\bfX_{\a}
=\bfx_{\a}-\bfx_{\b}$, as defined in
Fig.~\ref{objects}.  These formulas are known in the literature as
translation formulas.  A brief derivation of the translation
formulas can be found in Appendix \ref{app:C} and further discussion
in Ref.~\cite{ref:translation}.

Without loss of generality, to simplify the formulas we take
the Cartesian unit vectors that define the orientations of the
reference frames all to be parallel.  We will need only formulas that
apply to the outgoing solutions,
\begin{eqnarray}
\label{forms}
\cU^{\alpha\beta}_{l'm'lm}(\bfX_{\a\b}) &=& 
\sqrt{4\pi} (-1)^{m'} i^{l-l'}\sqrt{(2l+1)(2l'+1)} \cr
&& \times \sum_{l''m''} (-1)^{m''} i^{l''} \sqrt{2l''+1}
\begin{pmatrix}l&l'&l''\\0&0&0\end{pmatrix}
\begin{pmatrix}l&l'&l''\\m&-m'&-m''\end{pmatrix}
h_{l''}^{(1)}(kd_{\a\b})Y_{l''m''}(\hat\bfX_{\a\b})
\end{eqnarray}
The summation over $l''$ involves only a finite number of terms since
the 3-$j$ symbols vanish for $l''>l+l'$ and $l''<|l-l'|$.  Moreover
they are zero if $l+l'+l''$ is odd or if $m''\neq m-m'$.   The matrix
$\cU^{\alpha\beta}_{l'm'lm}$ has the following symmetries,
\begin{eqnarray}
\label{eq:V-symm-1}
\cU^{\alpha\beta}_{l'm'lm}&=&(-1)^{l+l'+m+m'} \cU^{\alpha\beta}_{lml'm'} \, ,\\
\label{eq:V-symm-2}
\cU^{\alpha\beta}_{l'm'lm}&=&\cU^{\alpha\beta}_{l'-m'l-m} \, \\
 {\cU^{\beta\alpha}_{l'm'lm}}&=&(-1)^{l+l'}\cU^{\alpha\beta}_{l'm'lm} \, .
\end{eqnarray}
If we are considering only two objects, then it is always possible to
orient the Cartesian coordinate systems so that they are related by
translation along the $z$-axis.  In this case
$\cU^{\alpha\beta}_{l'm'lm}$ simplifies. For $\bfX_{\alpha\beta}= \pm
d_{\alpha\beta} \hat \bfz$ one obtains, using
$Y_{l''m''}(\pm\hat\bfz)=(\pm)^{l''}\sqrt{(2l''+1)/4\pi}
\,\delta_{m''0}$,
\begin{equation}
  \label{eq:forms-along-z}
 \cU^{\alpha\beta}_{l'm'lm}( \pm
d_{\alpha\beta} \hat \bfz )= \delta_{m'm}
(-1)^{m} i^{l-l'}\sqrt{(2l+1)(2l'+1)}
\sum_{l''} i^{\pm l''} (2l''+1)
\begin{pmatrix}l&l'&l''\\0&0&0\end{pmatrix}
\begin{pmatrix}l&l'&l''\\m&-m&0\end{pmatrix} 
h_{l''}^{(1)}(kd_{\alpha\beta}) \, .
\end{equation}

\section{Evaluation of the Casimir Energy}
\label{eval}

\subsection{Performing the integral over $\phi$}
We start with the expression for the fixed-$k$ functional integral,
\Eq{source3}.  For any fixed sources, $\{\varrho_{\a}\}$, there is a
unique classical field, $\phicl[\varrho]$, that is the solution to
$\delta \widetilde S[\phi,\varrho]/\delta\phi(\bfx)=0$.  The classical
theory defined by $\widetilde S[\phi,\varrho]$, describes a complex
scalar field coupled to a set of sources on the surfaces, and is a
generalization of electrostatics.  By analogy with electrostatics, the
field $\phi$ is continuous throughout space, but its normal derivative
jumps by $\varrho_{\a}(\bfx )$ across $\Sigma_{a}$.  Indeed, the
classical equations of motion that follow from $\delta\widetilde
S/\delta\phi=0$ are
\begin{eqnarray}
\left(\nabla^{2} +k^{2}\right)\phicl(\bfx)&=& 0,
\quad\mbox{for $\bfx\notin \Sigma_\a$},\nonumber\\
\Delta \phicl(\bfx )&=& 0,
\quad\mbox{for $\bfx\in \Sigma_\a$,}\nonumber\\
\left.\Delta \p_{n}\phicl\right|_{\bfx }&=&\varrho_{\a}(\bfx ),
\quad \mbox{for $\bfx\in \Sigma_\a$,}
\label{variation}
\end{eqnarray}
where $\Delta\phi=\phi_{\rm in}-\phi_{\rm out}$ and
$\Delta\p_{n}\phi=\p_{n}\phi|_{\rm in}-\p_{n}\phi_{\rm out}$. The
subscripts ``in'' and ``out'' refer to the field inside and outside
the bounding surface $\Sigma_\a$.  As before, all normals point out of
the compact surfaces.  The solution to Eq.~\eqref{variation} is unique
up to solutions of the homogeneous equations, which we
exclude by demanding that $\phicl$ vanish when the
$\{\varrho_{\alpha}\}=0$.  Continuing the analogy with electrostatics,
we can write the classical field in terms of the free Green's function
and the sources,
\begin{equation}
\label{phib}
\phicl(\bfx) =\sum_{\beta}\int_{\Sigma_{\b}}d\bfx'
\cG_{0}(\bfx,\bfx',k)\varrho_{\b}(\bfx') \, ,
\end{equation}
where $\cG_{0}$ is the free Green's function given in Eq.~\eqref{gf3}. 

To compute the functional integral over $\phi$, {we} first
decompose $\phi$ into the classical part given by Eq.~\eqref{phib} and
a fluctuating part, 
\begin{equation}
\phi(\bfx) = \phicl(\bfx) +\delta \phi(\bfx) \, .
\end{equation}
Then, because the effective action, $\widetilde S$, is quadratic in
$\phi$, the $\delta\phi$ dependent terms are independent of $\phicl$,
\begin{equation}
\label{sourceintegral1}
\mathfrak{Z}_{\cC}(k)= \prod_{\a=1}^{N}
\int \left[{\cD}\varrho_{\a}(\bfx )\right]
e^{i {\frac{T}{\hbar}}
\widetilde S_{\rm cl}[\varrho]}\int [{\cD}\delta\phi (\bfx)]
\exp \left[i  {\frac{T}{\hbar}} \int
d\bfx\left(k^{2}|\delta\phi(\bfx)|^{2}- |\nabla\delta\phi(\bfx)|^{2}  
\right)\right] \, .
\end{equation}
The classical action can be simplified by using the equations of
motion, Eq.~\eqref{variation}, which make it possible to express the action
entirely in terms of integrals over the surfaces $\{\Sigma_{\a}\}$,
\begin{equation}
\label{classicalaction}
\widetilde S_{\rm cl}[\varrho] = {\frac{1}{2}}\sum_{\a}
\int_{\Sigma_{\a}} d\bfx \,
\left(\varrho^{*}_{\a}(\bfx )\phicl(\bfx )+\cc \right)\, ,
\end{equation}
where $\phicl(\bfx )$ is understood  to be a functional of the sources
{$\varrho_{\a}$}. 

The functional integral over $\delta\phi$ is \emph{independent of the
classical field $\phicl$} and defines the energy of the
unconstrained vacuum fluctuations of $\phi$.  This term is divergent, or,
more precisely, depends on some unspecified ultraviolet cutoff.
However it can be discarded because it is independent of the sources
and therefore common to $\mathfrak{Z}_{\cC}$ and $\mathfrak{Z}_{\infty}$.
Note that this result is an explicit demonstration of the contention of
Ref.~\cite{rlj}: the Casimir force has nothing to do with the
\emph{vacuum} fluctuations of $\phi$, but is instead a consequence of
the interaction between fluctuating sources in the materials.  It is
therefore not directly relevant to the fluctuations that are conjectured to be
associated with the dark energy.

From Eq.~\eqref{phib} it is clear that the solution to
Eq.~\eqref{variation} obeys the superposition principle:
$\phicl(\bfx)$ is a sum of contributions from each of the sources,
\begin{equation}
\phicl(\bfx)=\sum_{\beta}\phi_{\beta}(\bfx) \, ,
\end{equation}
where $\phi_{\beta}$ satisfies Eq.~\eqref{variation} with all sources
set equal to zero except for $\varrho_{\beta}$.  So the action can be
expressed as a double sum over surfaces and over contributions to
$\phicl$ generated by different objects.  This leaves a partition
function, $\mathfrak{Z}_{\cC}(k)$, of the form
\begin{equation}
\label{jintegral}
\mathfrak{Z}_{\cC}(k)= \prod_{\a=1}^{N}
\int \left[{\cD}\varrho_{\a}(\bfx )\right]
\exp\left[{\frac{i}{2} {\frac{T}{\hbar}} }\sum_{\a,\b}
\int_{\Sigma_\a} d\bfx \left(\varrho^{*}_{\a}(\bfx )
\phi_{\beta}(\bfx )+\cc\right)\right],
\end{equation}
to be evaluated.

\subsection{Evaluation of the Classical Action}

The classical action in Eq.~\eqref{jintegral} contains two
qualitatively different terms, the interaction between different
sources, $\a\ne\beta$, and the self-interaction of the source
$\varrho_{\a}$.  Both can be expressed as functions of the multipole
moments of the sources on the surfaces.

\subsubsection{Interaction terms:  $\a\ne\beta$}

Consider the contribution to the action from the field,
$\phi_{\beta}$, generated by the source, $\varrho_{\beta}$, integrated over
the surface $\Sigma_{\alpha}$,
\begin{equation}
\label{interaction}
\widetilde S_{\b\a}=\frac{1}{2}\int_{\Sigma_\a} d\bfx_\alpha  \left(
\varrho^{*}_{\a}(\bfx_\alpha )\phi_{\beta}(\bfx_\alpha)+\cc \right) \, ,
\end{equation}
where the subscript $\alpha$ on $\bfx_\alpha$ indicates that the
integration runs over coordinates measured relative to the origin of
object $\alpha$.  The field $\phi_{\b}(\bfx_\beta)$, \emph{measured
relative to the origin of object $\beta$}, can be represented as an
integral over its sources on the surface $\Sigma_{\b}$ as in
Eq.~\eqref{phib}.  Since every point on $\Sigma_{\alpha}$ is outside
of a sphere enclosing $\Sigma_{\b}$, the partial wave representation
of $\cG_{0}$ simplifies.  The coordinate $\bfx_{>}$ is always
associated with $\bfx_\beta$ and $\bfx_{<}$ is identified with
$\bfx'_\beta$, so Eq.~\eqref{phib} can be written
\begin{equation}
\label{phib2}
\phi_{\b}(\bfx_\beta)= {ik}\sum_{lm} h^{(1)}_{l}(k r_{\b})Y_{lm}(\hat\bfx_{\b})
\int_{\Sigma_{\b}}d\bfx'_{\b}j_{l}(k r'_{\b})
Y^{*}_{lm}(\hat\bfx'_{\b})\varrho_{\b}(\bfx'_{\b})\, .
\end{equation}
Note that the arguments of the Bessel functions and spherical
harmonics are all defined relative to the origin $\cO_{\b}$.  In
particular, $r'_{\b}$ and $\hat\bfx'_{\b}$ are the radial and angular
coordinates relative to $\cO_{\b}$ corresponding to a point
$\bfx'$ on the surface $\Sigma_{\b}$.  The integrals over
$\Sigma_{\b}$ define the \emph{multipole moments} of the source
$\varrho_{\b}$, which will be our final quantum variables,
\begin{equation}
\label{multipoles}
Q_{\b,lm}\equiv
\int_{\Sigma_{\b}}d\bfx_{\b}j_{l}(k r_{\b})
Y^{*}_{lm}(\hat\bfx_{\b})\varrho_{\b}(\bfx_{\b}),
\end{equation}
so that
\begin{equation}
\label{phib3}
\phi_{\b}(\bfx_\beta)= {ik}\sum_{lm}Q_{\b,lm} h^{(1)}_{l}(k r_{\b})
Y_{lm}(\hat\bfx_{\b}).
\end{equation}

The field $\phi_{\b}$ viewed from the surface $\Sigma_{\a}$ is a
superposition of solutions to the Helmholtz equation that are regular
at the origin $\cO_{\a}$.  Using the translation formulas,
Eq.~\eqref{eq:reg+out+solutions}, the
field generated by object $\Sigma_\beta$ can be written
as function of the coordinate $\bfx_\alpha$, measured from the
origin $\cO_\alpha$, as
\begin{equation}
\label{phib4}
\phi_{\b}(\bfx_\alpha)= {ik}\sum_{lm}Q_{\b,lm} \sum_{l'm'}
\cU^{\alpha\beta}_{l'm'lm}j_{l'}(kr_{\a})Y_{l'm'}(\hat\bfx_{\a}) \, .
\end{equation}
This result, in turn, can be substituted into the contribution $\widetilde
S_{\beta\alpha}$ to the action, leading to the simple result
\begin{equation}
\label{offdiag}
\widetilde S_{\b\a}[Q_{\a},Q_{\b}]=
{\frac{ik}{2}}\sum_{lml'm'}Q^{*}_{\a,l'm'}\cU^{\alpha\beta}_{l'm'lm}
Q_{\b,lm}+\cc \, .
\end{equation}
Note that the contributions to the action that couple fields and
sources on different objects make no reference to the particular
boundary conditions that characterize the Casimir problem.  They
depend only on the multipole moments of the fields and on the geometry
through the translation matrix $\mathbb{U}^{\a\b}$.

\subsubsection{Self-interaction terms}

We turn to the terms in $\widetilde S_\text{cl}$ where the field
and the source both refer to the same surface, $\Sigma_{\a}$:
\begin{equation}
\label{self1}
\widetilde S_{\a}[\varrho_{\a}]={\frac{1}{2}}
\int_{\Sigma_\a}  d\bfx \left( \varrho^{*}_{\a}(\bfx )
\phi_{\a}(\bfx) + \cc \right) \, .
\end{equation}
For the self-interactions terms, we only use the coordinate
system with origin $\cO_{\a}$ inside the surface
$\Sigma_\a$, and hence drop the label $\alpha$ on the coordinates
in this section.  Since $\phi_\alpha(\bfx)$ is continuous across the
surface, we can regard the $\phi_{\a}$ in Eq.~\eqref{self1} as the field
\emph{inside} $\Sigma_\a$, $\phi_{{\rm in},\a}$, which is a solution
to Helmholtz's equation that must be regular at the origin $\cO_{\a}$,
\begin{equation}
\label{self2}
\phi_{{\rm
in},\a}(\bfx)=\sum_{lm}\phi_{\a,lm}j_{l}(kr)Y_{lm}(\hat\bfx) 
\,  .
\end{equation}
Substituting this expansion into
Eq.~\eqref{self1}, we obtain
\begin{equation}
\label{self3}
\widetilde S_{\a}[\varrho_{\a}]=\frac{1}{2}
\sum_{lm}\left( \phi_{\a,lm}Q^{*}_{\a,lm}+\cc \right)\, ,
\end{equation}
where {the} $Q_{\a,lm}$ are the multipole moments of
the source{s,} defined in the previous subsection.

Finally we relate $\phi_{\a,lm}$ back to the multipole moments of the
source to get an action entirely in terms of the 
$Q_{\a{,lm}}$.  The field
$\phi_{\a,\rm{out}}$ at points \emph{outside} of $\Sigma_{\a}$ obeys
Helmholtz's equation and must equal $\phi_{\a,{\rm in}}$ on the
surface $S$.  Therefore it can be written as $\phi_{\a,{\rm in}}$
\emph{plus a superposition of the regular solutions to the Helmholtz
equation that vanish on $\Sigma_{\a}$}, as defined in Eq.~\eqref{prereg},
\begin{equation}
\label{outsidereg}
\phi_{\a,{\rm out}}(\bfx) =\phi_{\a,{\rm in}}(\bfx)+\Delta\phi_{\a}(\bfx)=
\phi_{\a,{\rm in}}(\bfx) +\sum_{lm}\chi_{\a,lm}\left(j_{l}(k r)
Y_{lm}(\hat\bfx)
+\sum_{l'm'}{\cal T}^\a_{l'm'l m }(k)h_{l'}^{(1)}(k r)
Y_{l'm'}(\hat\bfx)\right) \, .
\end{equation} 
The second term, $\Delta\phi_{\a}$, vanishes on $\Sigma_{\a}$ because
${\mathbb T}^{\a}$ is the scattering amplitude for the Dirichlet
problem.

The field we seek is generated in response to the sources and
therefore falls exponentially (for $k$ {with positive imaginary
  part}) as $r\to\infty$.  Therefore the terms in
Eq.~\eqref{outsidereg} that are proportional to $j_{l}(kr)$ must
cancel.  Comparing Eq.~\eqref{outsidereg} with Eq.~\eqref{self2}, we
conclude that $\chi_{\a,lm}=-\phi_{\a,lm}$, and therefore
\begin{equation}
\label{phiout}
\phi_{\a,{\rm out}}(\bfx) =-\sum_{lm}\phi_{\a,lm}\sum_{l'm'}
{\cal T}^{\a}_{l'm'l m }(k )h_{l'}^{(1)}(k r)Y_{l'm'}(\hat\bfx) \, .
\end{equation}
On the other hand, $\phi_{\a,{\rm out}}(\bfx)$ can be expressed as an
integral over the source as in Eq.~\eqref{phib},
\begin{equation}
\phi_{\a,{\rm out}}(\bfx)=\int_{\Sigma_{\a}} d\bfx' \cG_{0}(\bfx,\bfx'{,k})
\varrho_\alpha(\bfx')  \, .
\end{equation}
Using the partial wave expansion for the free Green's function,
Eq.~\eqref{gf3}, we find
\begin{equation}
\label{outagain}
\phi_{\a, {\rm out}}(\bfx)=ik\sum_{l'm'}Q_{\a,l'm'}h_{l'}^{(1)}(kr)Y_{l'm'}(\hat\bfx) ,
\end{equation}
and comparing with Eq.~\eqref{phiout}, we see that
\begin{equation}
ikQ_{\a,l'm'}=-\sum_{lm}\cT^{\a}_{l'm'lm}(k)\phi_{\a,lm}  ,\nonumber
\end{equation}
or 
\begin{equation}
\phi_{\a,lm}=-ik\sum_{l'm'}[\cT^{\a}]^{-1}_{lml'm'}Q_{\a,l'm'},
\end{equation}
where {$[\mathbb{T}^\alpha]^{-1}$} is the inverse of the Dirichlet
transition matrix $\mathbb{T}^\alpha$.  When this is combined with
Eq.~\eqref{self3}, we obtain the desired expression for the
self-interaction contribution to the action,
\begin{equation}
\label{selfaction}
\widetilde S_{\a}[Q_{\a}]={-\frac{ik}{2}}
\sum_{lml'm'}Q^{ *}_{\a,lm}[\cT^{\a}]^{-1}_{lml'm'}
Q_{\a,l'm'} + \cc \, . 
\end{equation}

\subsection{Evaluation of the Integral over Sources}

Combining Eq.~\eqref{selfaction} with Eq.~\eqref{offdiag}, we obtain
an expression for the action that is a quadratic functional of the
multipole moments of the sources on the surfaces.  The functional
integral Eq.~\eqref{jintegral} can be evaluated by changing variables
from the sources, $\{\varrho_{\a}\}$ to the multipole moments.  The functional
determinant that results from this change of variables can be
discarded because it is a common factor which cancels between
$\mathfrak{Z}_{\cC}$ and $\mathfrak{Z}_{\infty}$.  To
compute the functional integral we analytically continue to imaginary
frequency, $k=i\kappa$, $\kappa>0$,
\begin{equation}
\label{qintegral}
\mathfrak{Z}_{\cC}({i\kappa})=
\prod_{\a=1}^{N}\int \left[{\cD}Q_{\a}\cD Q^{*}_{\a}\right]
\exp\left\{{-\frac{\kappa}{2}}
{\frac{T}{\hbar}}
\sum_{\a}Q_{\a}^{*}[{\mathbb T}^{\a}]^{-1}Q_{\a}
+{\frac{\kappa}{2}}
{\frac{T}{\hbar}}
\sum_{\a\ne\b}Q_{\a}^{*}{\mathbb
U}^{\alpha\beta}Q_{\b}+ \cc \right\} \, ,
\end{equation}
where we have suppressed the partial wave indices.  The functional
integral Eq.~\eqref{qintegral} yields the inverse determinant of a matrix
$\mathbb{M}_{\cC}^{{\alpha\beta}}$ that is composed of the
inverse transition matrices $[{\mathbb T}^\alpha]^{-1}$ on its
diagonal and the translation matrices ${\mathbb U}^{\alpha\beta}$ on the
off-diagonals:
\begin{equation}
\mathbb{M}_{\cC}^{{\alpha\beta}} =
[\mathbb{T^{\alpha}}]^{-1}{\delta_{\alpha\beta}}
- \mathbb{U}^{{\alpha\beta}} {(1-\delta_{\alpha\beta})}\, ,
\end{equation}
  Finally we substitute into
Eq.~\eqref{casimir2} to obtain the Casimir energy,
\begin{equation}
\label{finalform}
\cE[\cC]=\frac{\hbar c}{\pi}\int_0^\infty d\kappa 
\ln\frac{\det\mathbb{M}_{\cC}(i\kappa)}{\det\mathbb{M}_{\infty}(i\kappa)} \, ,
\end{equation}
where the determinant is taken with respect to the partial wave
indices and the object indices $\alpha$, $\beta$, and
$\mathbb{M}_{\infty}^{\alpha\beta}=[\mathbb{T}^\alpha]^{-1}
\delta_{\alpha\beta}$
is the result of removing the objects to infinite separation, where
the interaction effects vanish.

In the special case of two interacting objects Eq.~\eqref{finalform}
simplifies to
\begin{equation}
\label{finalform2}
\cE_{2}[\cC]=\frac{\hbar c}{\pi}\int_0^\infty d\kappa 
\ln\det(1-{\mathbb T}^{1}\mathbb{U}^{12}\mathbb{T}^{2}\mathbb{U}^{21}) \, ,
\end{equation}
where ${\mathbb T}^{\a}, \a=1,2$, and $\mathbb{U}^{\a\b}$ are the
transition and translation matrices for the two objects. 

\begin{figure}
 \includegraphics[scale=0.7]{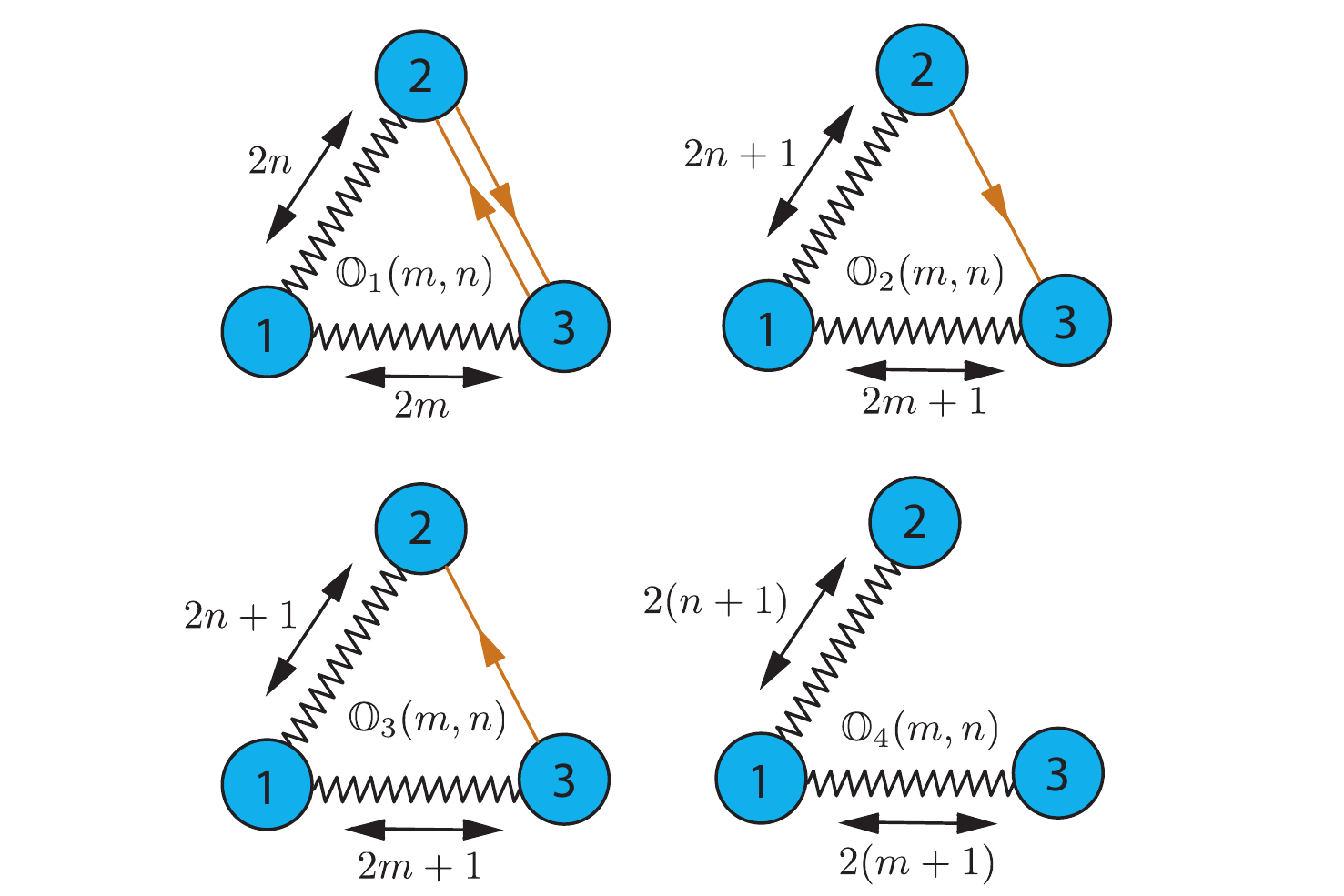}
\caption{(Color online) Scattering processes described by the operators of
  Eq.~\eqref{eq:4-scatt-operators}. The directed lines between objects
  $2$ and $3$ describe a wave travelling once between the objects. The
  zigzag lines correspond to multiple reflections in both directions
  where the number next to the line indicates how many times the wave
  travels between the objects. To interpret the diagrams one starts
  with a wave at object $3$. As an example, we describe explicitly the
  diagram for the operator ${\mathbb O}_1(m,n)$. The wave travels from
  object $3$ to object $2$, is then reflected $2n$ times back and
  forth between objects $2$ and $1$, travels back from object $2$ to
  $3$, and is finally reflected $2m$ times back and forth between
  objects $3$ and $1$. Each free propagation between the objects is
  described by the translation operator ${\mathbb U}^{\alpha\beta}$,
  and each reflection at an object by the transition operator
  ${\mathbb T}^\alpha$.}
\label{3-body-scatt}
\end{figure}

For three
objects $\cE$ takes the form,
\begin{eqnarray}
\label{finalform3}
\cE_{3}[\cC]&=&\frac{\hbar c}{\pi}\int_0^\infty d\kappa \left\{ \ln\det(
1-{\mathbb T}^{2}\mathbb{U}^{21}\mathbb{T}^{1}\mathbb{U}^{12}) 
+ \ln\det(
1-{\mathbb T}^{3}\mathbb{U}^{31}\mathbb{T}^{1}\mathbb{U}^{13})
\right. \nonumber\\
&+&\left.\ln\det\left[1-(1-{\mathbb T}^{3}{\mathbb U}^{31}
{\mathbb T}^{1}{\mathbb U}^{13})^{-1}
({\mathbb T}^{3}{\mathbb U}^{32}+{\mathbb T}^{3}{\mathbb U}^{31}
{\mathbb T}^{1}{\mathbb U}^{12})
(1-{\mathbb T}^{2}{\mathbb U}^{21}{\mathbb T}^{1}{\mathbb U}^{12})^{-1}
({\mathbb T}^{2}{\mathbb U}^{23}+{\mathbb T}^{2}{\mathbb U}^{21}
{\mathbb T}^{1}{\mathbb U}^{13})
\right]
\right\}\, .\nonumber\\
\end{eqnarray}
Although this result appears complicated, it admits a simple physical
interpretation. The first two terms describe scatterings between
objects $1$ and $2$ and objects $1$ and $3$, and hence correspond to
the separate two-body Casimir energies of these two pairs of
objects. The third term must have a more complicated form because
fluctuation forces are not pairwise additive and hence the third term
is {\it not} simply given by the two-body Casimir energy of objects
$2$ and $3$. The interaction of objects $2$ and $3$ involves not only
direct scatterings between these two objects but also indirect
multiple scatterings off object $1$.  The third term in
Eq.~\eqref{finalform3} contains the resolvent of the operator
${\mathbb T}^{3}{\mathbb U}^{31}{\mathbb T}^{1}{\mathbb U}^{13}$,
which can be formally expanded as the series $\sum_{n=0}^\infty
({\mathbb T}^{3}{\mathbb U}^{31}{\mathbb T}^{1}{\mathbb
  U}^{13})^n$.  The resolvent of ${\mathbb T}^{2}{\mathbb
  U}^{21}{\mathbb T}^{1}{\mathbb U}^{12}$ can be treated
  similarly. These series describe multiple scatterings between
objects $1$ and $3$, and objects $1$ and $2$, respectively,
which occur as intermediate steps in the scattering processes
between objects $2$ and $3$. Inserting the series into the last term
of Eq.~\eqref{finalform3}, one can distinguish four
qualitative{ly} different scattering processes, described by the
following operators,
\begin{eqnarray}
\label{eq:4-scatt-operators}
{\mathbb O}_1(m,n) &=& \left({\mathbb T}^{3}{\mathbb U}^{31}
{\mathbb T}^{1}{\mathbb U}^{13}\right)^m {\mathbb T}^3 {\mathbb U}^{32} 
\left({\mathbb T}^{2}{\mathbb U}^{21}
{\mathbb T}^{1}{\mathbb U}^{12}\right)^n 
{\mathbb T}^2 {\mathbb U}^{23}\nonumber\\
{\mathbb O}_2(m,n) &=& \left({\mathbb T}^{3}{\mathbb U}^{31}
{\mathbb T}^{1}{\mathbb U}^{13}\right)^m {\mathbb T}^3 {\mathbb U}^{32} 
\left({\mathbb T}^{2}{\mathbb U}^{21}{\mathbb T}^{1}
{\mathbb U}^{12}\right)^n
{\mathbb T}^2 {\mathbb U}^{21} {\mathbb T}^1 {\mathbb U}^{13}
\nonumber\\
{\mathbb O}_3(m,n) &=& \left({\mathbb T}^{3}{\mathbb U}^{31}
{\mathbb T}^{1}{\mathbb U}^{13}\right)^m {\mathbb T}^3
{\mathbb U}^{31} {\mathbb T}^1 {\mathbb U}^{12} 
\left({\mathbb T}^{2}{\mathbb U}^{21}{\mathbb T}^{1}
{\mathbb U}^{12}\right)^n  {\mathbb T}^2 {\mathbb U}^{23}
\nonumber\\
{\mathbb O}_4(m,n) &=& \left({\mathbb T}^{3}{\mathbb U}^{31}
{\mathbb T}^{1}{\mathbb U}^{13}\right)^m {\mathbb T}^3 
{\mathbb U}^{31} {\mathbb T}^1 {\mathbb U}^{12} 
\left({\mathbb T}^{2}{\mathbb U}^{21}{\mathbb T}^{1}
{\mathbb U}^{12}\right)^n  {\mathbb T}^2 {\mathbb U}^{21}
{\mathbb T}^1 {\mathbb U}^{13}
\, .
\end{eqnarray}
The operator ${\mathbb O}_1(0,0)= {\mathbb T}^3 {\mathbb U}^{32}
{\mathbb T}^2 {\mathbb U}^{23}$ describes direct scatterings between
objects $2$ and $3$ as if object $1$ were absent,
and hence yields the additive approximation for the energy. All
additional operators of Eq.~\eqref{eq:4-scatt-operators} describe the
non-additivity of the Casimir interaction between three objects. The
corresponding scattering processes are depicted in
Fig.~\ref{3-body-scatt}. The operator ${\mathbb O}_1(m,n)$ describes
two scatterings between objects $2$ and $3$ with $m$ ($n$) scatterings
between objects $3$ ($2$) and $1$ in between. The operators
${\mathbb O}_2(m,n)$ (${\mathbb O}_3(m,n)$) correspond to a wave that
travels from object $3$ ($2$) to object $2$ ($3$) and is reflected
back to object $3$ ($2$) from object $1$ after multiple scatterings
between objects $1$ and $2$, and $1$ and $3$, respectively. Finally,
operator ${\mathbb O}_4(m,n)$ describes only indirect scattering
processes between objects $2$ and $3$ that all go through object $1$.

\section{Applications}
\label{appl}

In this section we give a few typical applications of our method.
Although the physically interesting case of electromagnetism requires
vector fields and conducting boundary conditions, the scalar case
offers an opportunity to check our methods and illustrate their
utility.  We consider a {\it real} scalar field fluctuating in the
space between two spheres on which Robin boundary conditions, $\phi -
\lambda_\alpha \partial_n \phi=0$, are imposed.  Because a real field
has half the oscillation modes of a complex field, the Casimir energy
in Eq.~(\ref{finalform2}) must be divided by 2, giving
\begin{equation}
\label{finalform2real}
\cE_{2}[\cC]=\frac{\hbar c}{2\pi}\int_0^\infty d\kappa 
\ln\det(1-{\mathbb T}^{1}\mathbb{U}^{12}\mathbb{T}^{2}\mathbb{U}^{21}) \, .
\end{equation}
We allow for different Robin parameters $\lambda_{1,2}$ and different
radii $R_{1,2}$ for the two spheres.  This choice allows us to study
Dirichlet ($\lambda/R\to 0$) and Neumann ($\lambda/R\to\infty$)
boundary conditions on separate spheres as special cases. We obtain
the Casimir energy as a series in $R_{1}/d$ and $R_{2}/d$ for large
separations $d$ and numerically at all separations.  A comparison of
the two approaches allows us to measure the rate of convergence of our
results.  We find that for Robin boundary conditions the sign of the
force depends on the ratios $\lambda_{\a}/R_{\a}$ and on the
separation $d$.  We also express the Casimir energy between two
objects of general shape in a large distance expansion in terms of the
coefficients of capacitance defined in Sec.~\ref{2.3}.

\subsection{Interaction of two spheres with Robin boundary conditions:  
general considerations}

The Robin boundary condition $\phi - \lambda_\alpha \partial_n \phi=0$
allows a continuous interpolation between Dirichlet and Neumann
boundary conditions. Since the radius of the sphere introduces a
natural length scale, it is convenient to replace $\lambda$
by a dimensionless variable, $\zeta{_{\a}}\equiv
\lambda{_{\a}}/R{_{\a}}$.  For
$\zeta_\alpha>0$, the modulus of the field is suppressed if the
surface is approached from the outside, while for $\zeta_\alpha<0$ it
is enhanced. Hence, for negative $\zeta_\alpha$ bound surface states
can be expected.  All the information about the shape of the
object and the boundary conditions at its surface is provided by the
$\mathbb{T}$-matrix.  For spherically symmetric objects the
$\mathbb{T}$-matrix is diagonal and is completely specified by phase
shifts $\delta_{l}{(k)}$ that do not depend on $m$,
\begin{equation}
  \label{eq:spheres-t-from-shifts}
  \cT_{lml'm'}(k)=\delta_{ll'}\delta_{mm'} \frac{1}{2} 
\left( e^{2i\delta_l(k)}-1\right) \, . 
\end{equation}
In the discussion of the $\mathbb{T}$-matrix for an individual
object we {again} suppress the label $\alpha$.  The phase shifts
for Robin boundary conditions are
\begin{equation}
  \label{eq:sphere-phases}
  \cot \delta_l(k) = \frac{n_l(\xi)-\zeta\xi\, n'_l(\xi)}
{j_l(\xi)-\zeta\xi\, j'_l(\xi)} \, ,
\end{equation}
where $\xi=kR$ and $j_l~(n_l)$ are spherical Bessel functions of
first(second) kind. To apply Eq.~\eqref{finalform2real}, we have to
evaluate the matrix elements of the transition matrices for imaginary
frequencies $k=i\kappa$.  Using  $j_l(iz)=i^l
\sqrt{\pi/(2z)} I_{l+1/2}(z)$ and $h^{(1)}_l(iz)= -i^{-l} \sqrt{2/(\pi
  z)} K_{l+1/2}(z)$, we obtain for the $\mathbb{T}$-matrix elements
\begin{equation}
  \label{eq:spheres-t-matrix-imag}
  \cT_{lmlm}(i\kappa)=(-1)^l \frac{\pi}{2} \frac{(1/\zeta+1/2) I_{l+1/2}(z)
-z I'_{l+1/2}(z)}{(1/\zeta+1/2) K_{l+1/2}(z)
-z K'_{l+1/2}(z)} \, ,
\end{equation}
where $z\equiv \kappa R$.

For two spherical objects we can assume that the center-to-center
distance vector is parallel to the $z$-axis. Then the translation
matrices simplify to the form presented in
Eq.~\eqref{eq:forms-along-z}. For imaginary frequencies the
translation matrix elements become
\begin{equation}
  \label{eq:spheres-transl-matrix-imag}
\cU^{\left\{\begin{matrix}12\\ 21\end{matrix}\right\}}_{l'mlm}(d)= 
- (-1)^{m} i^{-l'+l} \sqrt{(2l+1)(2l'+1)}\, \sum_{l''} \,
 (\pm 1)^{l''} (2l''+1)
\begin{pmatrix}l&l'&l''\\0&0&0\end{pmatrix}
\begin{pmatrix}l&l'&l''\\m&-m&0\end{pmatrix}
K_{l''+1/2}(\kappa d) \, ,
\end{equation}
where $d$ is the separation distance.

For a range of $\zeta$, bound states can appear.  This range can
be determined by examining the $\mathbb{T}$-matrix at small
frequencies, where it has the leading form
\begin{equation}
  \label{eq:sphere-T-small-k}
  \cT_{lmlm} \sim (-1)^l\frac{1-l\zeta}{1+(l+1) \zeta} 
z^{2l+1} +\ldots \, ,
\end{equation}
where we have suppressed positive numerical coefficients. The
amplitudes diverge for $\zeta=\lambda/R=-1/(l+1)$ indicating the
existence of a pole at $\kappa=0$ at this value of $\zeta$.  Indeed,
from Eq.~\eqref{eq:spheres-t-matrix-imag} we find that the $\mathbb{T}$-matrix
elements have a pole in the $l^\text{th}$ partial wave at $\kappa_l>0$
if $-1/(l+1)< \zeta <0$.  The values of $\kappa_{l}$ are determined by
the equation
\begin{equation}
  \label{eq:sphere-T-poles}
  - \frac{1}{\zeta} =  1+l+ R\kappa_l 
\frac{K_{l-1/2}(R\kappa_l)}{K_{l+1/2}(R\kappa_l)} \, .
\end{equation}
These poles correspond to bound states for $-1 < \zeta <0$. For any
$\zeta$ in this interval, there exists a finite number of bound
states, which increases as $\zeta \to 0$. In the following, we
restrict to $\zeta\ge 0$ and leave the study of interactions in the
presence of bound states to a future publication.

The special case of spheres with Dirichlet boundary conditions has
been studied in Ref.~\cite{Bulgac:2005ku}. For two spheres of equal
radius, the matrix $\sum_{l''}A_{ll''}^{(m)}A_{l''l'}^{(m)}$ in the
notation of Ref.~\cite{Bulgac:2005ku} is proportional to our
$\mathbb{T}^{1}\mathbb{U}^{12}\mathbb{T}^{2}\mathbb{U}^{21}$ times
$K_{l+1/2}(\kappa R)/K_{l'+1/2}(\kappa R)$. It is easy to see that
this proportionality factor drops out in the final result for the
energy if one uses $\ln \det = \tr \ln$ in Eq.~\eqref{finalform2real}
and expands the logarithm around unity. Thus we agree with the results
given in Ref.~\cite{Bulgac:2005ku}.

\subsection{Asymptotic expansion for large separation}

In this section we consider the Casimir interaction between two
spheres due to a scalar field obeying Robin boundary conditions,
allowing for a different parameter $\lambda_{1,2}$ on each sphere.
The Casimir energy can be developed in an asymptotic expansion in
$R_\alpha/d$ using $\ln \det = \tr \ln$ in Eq.~\eqref{finalform2real}.
Expanding the logarithm in powers of $\mathbb{N}= \mathbb{T}^{1}
\mathbb{U}^{12} \mathbb{T}^{2} \mathbb{U}^{21}$, since the
$\mathbb{T}$-matrix has no poles in the region of interest
we get
\begin{equation}
  \label{eq:spheres-energy-real}
  \cE = - \frac{\hbar c}{2\pi} \int_0^\infty d\kappa \sum_{p=1}^\infty 
\frac{1}{p} \tr\left(\mathbb{N}^p\right) \, .
\end{equation}
We have performed the matrix operations using {\tt Mathematica}. The
scaling of the $\mathbb{T}$-matrix at small $\kappa$ shows that the 
$p^\text{th}$ power of $\mathbb{N}$ (corresponding to $2p$
scatterings) become{s} important at order $d^{-(2p+1)}$. Partial waves of
order $l$ start to contribute at order $d^{-(3+2l)}$ if the
$\mathbb{T}$-matrix is diagonal in $l${,} which is the case for
spherically symmetric objects. Hence the leading terms with $p=1$ and
$l=0$ yield the exact energy to order $d^{-4}$. In the following we
will usually restrict the expansion to $p \le 3$, $l \le 2$,
yielding the interaction to order $d^{-8}$.

\subsubsection{Equal radii}
We begin with the case $R_{1}=R_{2}=R$.  The large distance expansion
of the Casimir energy can be written as
\begin{equation}
\label{eq:spheres-energy-large-d}
\cE = \frac{\hbar c }{\pi} \frac{1}{d}
\sum_{j=3}^\infty b_j \left(\frac{R}{d}\right)^{j-1} \, ,
\end{equation}
where $b_j$ is the coefficient of the term $\sim d^{-j}$.  For general
Robin boundary conditions we find
\begin{eqnarray}
\label{eq:sphere-energy-coeff}
b_3 &=& - \frac{1}{4 (1+\zeta_1)(1+\zeta_2)}, \\
b_4 &=& -\frac{ (2+\zeta_1+\zeta_2)}{8(1+\zeta_1)^2(1+\zeta_2)^2}, \\
b_5 &=& \Big[ -77 -231 (\zeta_1+\zeta_2) 
-\left(188 (\zeta_1^2+\zeta_2^2)+625 \zeta_1 \zeta_2\right) 
-2 \left(3 (\zeta_1^3+\zeta_2^3)+197 (\zeta_1 \zeta_2^2+ \zeta_2 \zeta_1^2)
\right) \nonumber\\
&+&2 \left(17 (\zeta_1^4+\zeta_2^4)+56 (\zeta_2 \zeta_1^3
+ \zeta_1 \zeta_2^3 )-18 \zeta_1^2 \zeta_2^2\right) 
+2 \zeta_1 \zeta_2 \left(68 (\zeta_1^3+\zeta_2^3)+155( \zeta_1
\zeta_2^2 +\zeta_2\zeta_1^2)\right) \nonumber\\
&+&2 \zeta_1^2 \zeta_2^2 \left(85 (\zeta_1^2+\zeta_2^2)
+124 \zeta_2 \zeta_1\right) 
+68 \zeta_1^3\zeta_2^3 (\zeta_1+\zeta_2)\Big]\,
\left[48 (1+\zeta_1)^3(1+2 \zeta_1) (1+\zeta_2)^3 
(1+2 \zeta_2)\right]^{-1},\nonumber\\
b_6&=&  \Big[ -50  -175  (\zeta_1+\zeta_2) 
- \left( 179(\zeta_1^2+\zeta_2^2)+592 \zeta_1\zeta_2 \right) 
- \left( 18 (\zeta_1^3+\zeta_2^3)
+565 (\zeta_1^2\zeta_2 + \zeta_2^2\zeta_1) \right) \nonumber\\
&+& \left( 64(\zeta_1^4+\zeta_2^4) +6(\zeta_1^3\zeta_2+\zeta_1\zeta_2^3) 
-460 \zeta_1^2\zeta_2^2\right) 
+ 2 \left( 11(\zeta_1^5+\zeta_2^5)+136(\zeta_1^4\zeta_2+\zeta_1\zeta_2^4)
+67(\zeta_1^3\zeta_2^2+\zeta_1^2\zeta_2^3) \right) \nonumber\\
&+&4\zeta_1\zeta_2 \left(22(\zeta_1^4+\zeta_2^4)
+93(\zeta_1^3\zeta_2+\zeta_1\zeta_2^3)
+52\zeta_1^2\zeta_2^2 \right) 
+2 \zeta_1^2\zeta_2^2\left(55(\zeta_1^3+\zeta_2^3)+
92(\zeta_1^2\zeta_2+\zeta_1\zeta_2^2) \right) \nonumber\\
&+&4 \zeta_1^3 \zeta_2^3 \left( 11 (\zeta_1^2+\zeta_2^2) 
+8 \zeta_1\zeta_2\right)
\Big]\, \left[ 32 (1+\zeta_1)^4(1+2 \zeta_1) (1+\zeta_2)^4
(1+2 \zeta_2)\right]^{-1} \, .
\end{eqnarray}
Higher order coefficients can be obtained but are not shown here in
order to save space. For special values of $\lambda_\alpha$ they are
given below. These coefficients show some interesting properties. They
all diverge for $\zeta_{\a}=-1$, where $\lambda_\alpha=-R_\alpha$, and
$b_5$ and $b_6$ also diverge for $\zeta_\alpha=-1/2$.  These results
are consistent with the emergence of poles in the $\mathbb{T}$-matrix
at small $\kappa$ when $\zeta_\alpha$ approaches $-1/(l+1)$ and the
observation that the partial wave with $l=1$ starts to contribute only
at order $d^{-5}$.  Another interesting
property is that some coefficients $b_j$
go to zero for $\lambda_\alpha\to\infty$, which
corresponds to Neumann boundary conditions. If both
$\lambda_\alpha$ go to infinity, the
coefficients $b_j$ vanish for $j=1,\ldots, 6$, so
that the leading term in the Casimir energy is $\sim d^{-7}$ with a
negative amplitude. Hence, Neumann boundary conditions lead to an
attractive Casimir-Polder power law, as is known from electromagnetic
field fluctuations.  This result can be understood from the absence of
low-frequency $s$-waves for Neumann boundary conditions. It is clearly
reflected by the low frequency expansion of
Eq.~(\ref{eq:sphere-T-small-k}){,} which has a vanishing amplitude
for $\lambda_\alpha\to\infty$ if $l=0$.  If one $\lambda_\alpha$
remains finite and the other goes to infinity, only
the coefficients $b_3$ and $b_4$ vanish so that the energy scales as
$d^{-5}$ with a {\it positive} amplitude.  Since $b_3<0$ for
$\lambda_\alpha\ge 0$, at asymptotic distances the Casimir force is
therefore attractive for all non-negative finite $\lambda_\alpha$, and
for $\lambda_\alpha$ both infinite.  It is repulsive if one
$\lambda_\alpha$ is finite and the other infinite, i.e., if one sphere
obeys Neumann boundary conditions. However, at smaller distances the
interaction can change sign depending on $\lambda_\alpha$, as shown
below.

More precisely, one has the following limiting cases. If both
$\lambda_\alpha=0$, the field obeys Dirichlet conditions at the two
spheres and the first six coefficients are
\begin{equation}
\label{eq:coeff_D-D}
b_3=-\frac{1}{4}, \quad b_4=-\frac{1}{4}, \quad b_5=-\frac{77}{48}, \quad 
b_6=-\frac{25}{16}, \quad 
b_7=-\frac{29837}{2880}, \quad b_8=-\frac{6491}{1152} \, .
\end{equation}
If Neumann conditions are imposed on both surfaces, the coefficients are
\begin{equation}
\label{eq:coeff_N-N}
b_3=0, \quad b_4=0, \quad b_5=0, \quad 
b_6=0, \quad b_7=-\frac{161}{96}, \quad b_8=0, \quad b_9=-\frac{3011}{192}
\quad b_{10}=-\frac{175}{128}\, ,
\end{equation}
clearly showing that the asymptotic interaction has a Casimir-Polder
power law $\sim {{\cal O}(}d^{-7}{)}$. Also, as in the
electromagnetic case, the next to leading order ${\cal O}(d^{-8})$
vanishes \cite{Emig:2007cf}.  Therefore we have
included the two next terms of the series.  If $\lambda_2\to\infty$
(Neumann conditions) and $\lambda_1$ remains finite, we have
\begin{eqnarray}
  \label{eq:coeff_N-finite}
  b_3&=&0, \quad b_4=0, \quad
  b_5=\frac{17  }{48(1+\zeta_1)}, \quad b_6=\frac{11 }{32(1+\zeta_1)^2} ,
\nonumber\\
  b_7&=&\frac{1989+4736 \zeta_1+1895  \zeta_1^2 -2192  \zeta_1^3
-1610 \zeta_1^4}{480(1+\zeta_1)^3(1 +2\zeta_1)} ,
\nonumber\\
  b_8&=&\frac{5(94+143 \zeta_1 +76 \zeta_1^2 -45 \zeta_1^3)}
{288 (1+\zeta_1)^4} \, .
\end{eqnarray}
Thus, for mixed Dirichlet/Neumann boundary conditions, the result is
\begin{equation}
  \label{eq:coeff_D-N}
    b_3=0, \quad b_4=0, \quad b_5=\frac{17}{48}, \quad 
b_6=\frac{11}{32}, \quad b_7=\frac{663}{160}, \quad b_8=\frac{235}{144} \, .
\end{equation}
Finally, if $\lambda_2=0$ (Dirichlet conditions) and $\lambda_1$ is finite, we
obtain
\begin{eqnarray}
\label{eq:coeff_D-finite}
b_3&=&-\frac{1}{4(1+\zeta_1)}, \quad 
b_4=-\frac{(2+\zeta_1)}{8(1+\zeta_1)^2}, 
\nonumber\\
b_5&=&-\frac{77 +231  \zeta_1+188  \zeta_1^2 +6 \zeta_1^3
-34 \zeta_1^4}{48(1+\zeta_1)^3(1+2\zeta_1)} ,
\nonumber\\
b_6&=&-\frac{50+175 \zeta_1 +179 \zeta_1^2 +18  \zeta_1^3
-64  \zeta_1^4 -22 \zeta_1^5}
{32 (1+\zeta_1)^4(1+2\zeta_1)} \, ,
\end{eqnarray}
where we have not shown higher order coefficients. It is important to
note that the series in Eq.~(\ref{eq:spheres-energy-large-d}) is an
asymptotic series and therefore cannot be used to obtain the
interaction at short distances.

\subsubsection{Unequal radii}
To study the individual contributions from two objects to the various
terms in the large distance expansion, it is instructive
to study two spheres of different radii $R_1$ and $R_2$. For
simplicity, we focus on Dirichlet and Neumann boundary conditions. By
expanding the energy in powers of $\mathbb{N}$ as before, we get an
asymptotic expansion of the Casimir energy,
\begin{equation}
  \label{eq:expansion-different-r}
\cE = \frac{\hbar c }{\pi} \frac{1}{d} \sum_{j=3}^\infty \tilde b_j(\eta) 
\left(\frac{R_1}{d}\right)^{j-1} \, ,  
\end{equation}
where the coefficients now also depend on $\eta=R_2/R_1$. 
For Dirichlet boundary conditions on both spheres{,} the coefficients are
\begin{eqnarray}
\label{eq:coeff_diff_r_DD}
\tilde b_3 &=&  -\frac{\eta}{4}, \quad \tilde b_4 =
-\frac{\eta+\eta^2}{8}, \quad
\tilde b_5 = -\frac{34 (\eta+\eta^3) + 9 \eta^2}{48}, \quad 
\tilde b_6 = -\frac{2 (\eta+\eta^4)+23(\eta^2+\eta^3)}{32} ,\\
\tilde b_7 &=& -\frac{8352 (\eta+\eta^5)+1995(\eta^2+\eta^4)
+38980 \eta^3}{5760}, \quad
\tilde b_8 = -\frac{-1344(\eta+\eta^6)+5478(\eta^2+\eta^5)
+2357(\eta^3+\eta^4)}{2304}
\, ,\nonumber
\end{eqnarray}
while for Neumann boundary conditions on both spheres{,} we have
\begin{equation}
  \label{eq:coeff_diff_r_NN}
  \tilde b_7 = -\frac{161 \eta^3}{96}, \quad \tilde b_8 = 0, \quad 
\tilde b_9 = -\frac{3011(\eta^3+\eta^5)}{384}, \quad
\tilde b_{10} = -\frac{175(\eta^3+\eta^6)}{256}
\, ,
\end{equation}
and for Dirichlet conditions on the sphere with radius
$R_1$ and Neumann conditions on the one with radius
$R_2${,} we obtain
\begin{equation}
\label{eq:coeff_diff_r_DN}
\tilde b_5 = \frac{17\eta^3}{48}, \quad 
\tilde b_6 = \frac{11\eta^3}{32}, \quad
\tilde b_7 = \frac{\eta^3(1610+379\eta^2)}{480}, \quad \tilde b_8 = 
\frac{5\eta^3(24+67\eta^2+3\eta^3)}{288}
\, .
\end{equation}
For like boundary conditions, the coefficients satisfy the symmetry
condition $\eta^{j-1}\tilde b_j(1/\eta) = \tilde b_j(\eta)$. The
coefficient $\tilde b_8$ for Dirichlet conditions becomes positive if
$\eta$ is sufficiently large or small compared to unity.  Due to the
absence of monopoles, Neumann conditions on one or both spheres lead
to coefficients that scale at least as $\eta^3$.  For the same reason,
not all powers of $\eta$ contribute to the coefficients if Neumann
conditions are imposed.  Note that the interaction between a sphere
and a plate cannot be obtained from this asymptotic expansion as the
limit $\eta\to 0$ since the expansion assumes that $R_1$, $R_2 \ll
d$. However, the sphere-plate interaction can be computed by the same
technique employed here if it is applied to the Green{'s} function of the
semi-infinite space bounded by the plate instead of the free Green{'s}
function.

\subsection{Numerical results for Robin boundary conditions on two spheres 
at all separations}

The primary application of our analysis is to compute the Casimir
energy and force to high accuracy over a broad range of distances.
However, to obtain the interaction at all distances,
Eq.~\eqref{finalform2real} has to be evaluated numerically.  We shall
see that the domain where our method is \emph{least} accurate is when
the two surfaces approach one another.  That is the regime where
semiclassical methods like the proximity force approximation (PFA)
become exact.  Because of its role in this limit, and because it is
often used (with little justification) over wide ranges of
separations, it is important to compare our calculations with the PFA
predictions.

\subsubsection{PFA for Robin boundary conditions}

\begin{figure}
\includegraphics[scale=0.4]{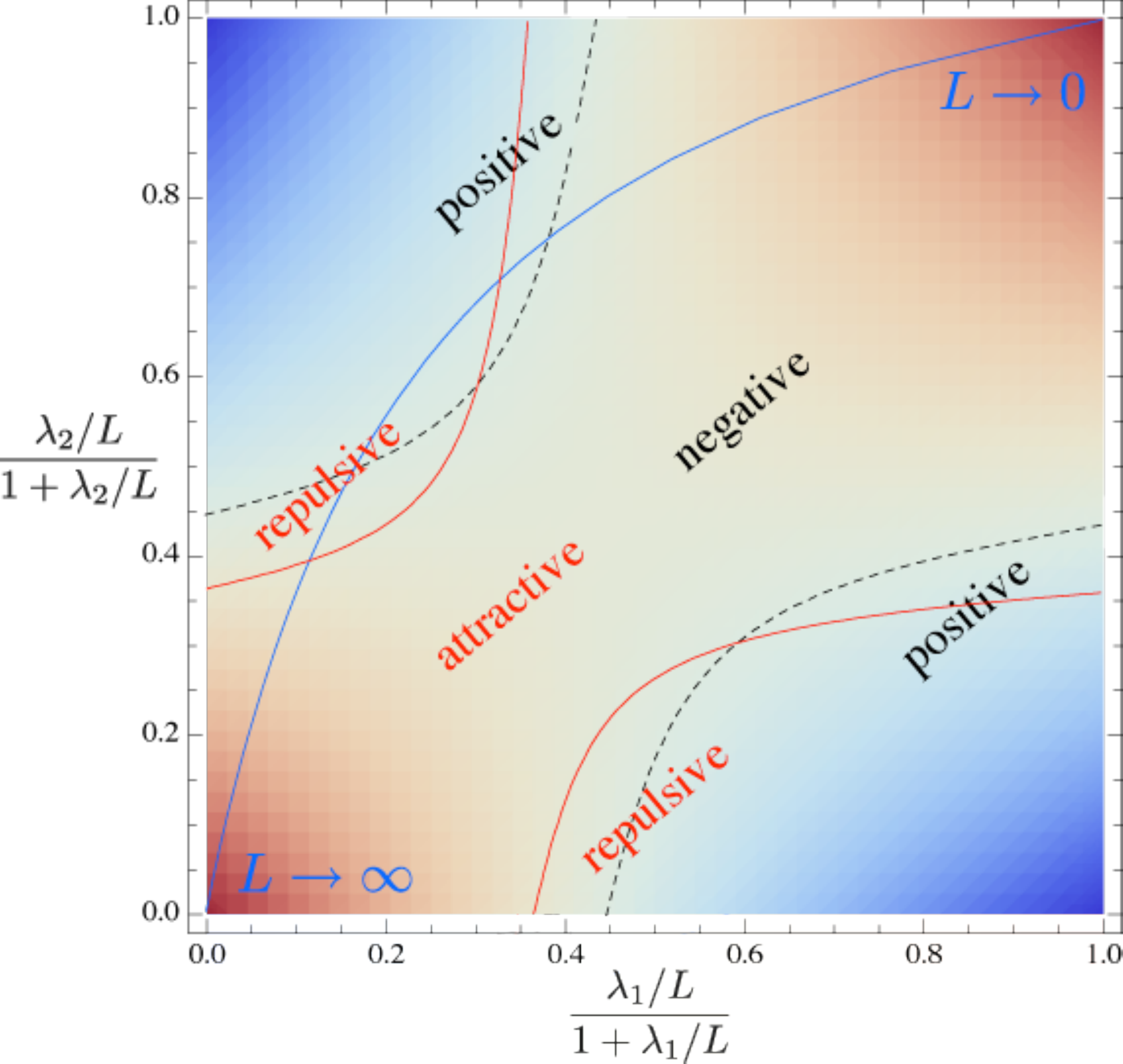}
\caption{(Color online) Amplitudes $\Phi(\lambda_1/L,\lambda_2/L)$ of
  the Casimir energy (from \eq{eq:Energy-Robin-pp}) and
  $\Phi'(\lambda_1/L,\lambda_2/L)$ of the Casimir force (from
  \eq{eq:Force-Robin-pp}) for two parallel plates of distance $L$ with
  Robin boundary conditions.  $L=0$ corresponds to the point $(1,1)$
  unless one or both plates have Dirichlet ($\lambda_\alpha=0$)
  boundary conditions. The color coding corresponds to the magnitude
  of the energy amplitude with blue corresponding to positive energy
  and red to negative.  The dashed (black) curves correspond to zero
  energy while the solid (red) curves represent zero force.  The curve
  running from the origin to $(1,1)$ shows as an example the
  interaction energy for $\lambda_2/\lambda_1=5$ as function of the
  distance $L$. For finite $\lambda_\alpha$ the interaction is always
  attractive at asymptotically large and small distances but can be
  repulsive at intermediate distances if the ratio
  $\lambda_1/\lambda_2 \gtrsim 2.8$ (or $\lambda_2/\lambda_1 \gtrsim
  2.8$). For $\lambda_1=\lambda_2$ the interaction is always
  attractive. The amplitudes are independent of $L$ for the special
  cases where the $\lambda_\alpha$ are either zero or infinite: The
  interaction is attractive (negative energy) for
  $\lambda_1=\lambda_2=0$, $\lambda_1=\lambda_2=\infty$ and repulsive
  (positive energy) for $\lambda_1=0$, $\lambda_2=\infty$ and
  $\lambda_1=\infty$, $\lambda_2=0$.}
\label{fig:plates-amplitude}
\end{figure}

In the proximity force approximation, the energy is obtained as an
integral over infinitesimal parallel surface elements at their local
distance $L$, measured perpendicular to a surface $\Sigma$
that can be one of the two surfaces of the objects, or an auxiliary
surface placed between the objects. The PFA approximation for the
energy is then given by
\begin{equation}
  \label{eq:energy-pfa}
  \cE_\text{PFA}=\frac{1}{A}\int_\Sigma \cE_\|(L) dS \, ,
\end{equation}
where $\cE_\|(L)/A$ is the energy per area for two parallel plates
with distance $L$. Functional integral techniques can be used to
compute the energy for parallel plates with mixed (Robin) boundary
conditions~\cite{LK91}. For brevity we only quote the results
\cite{Romeo+02}.  The energy is
\begin{equation}
\label{eq:Energy-Robin-pp}
\cE_\|(L) = \frac{\hbar c}{L^3} \Phi(\lambda_1/L,\lambda_2/L), \quad
\Phi(\lambda_1/L,\lambda_2/L)= \frac{1}{4\pi^2}\int_0^\infty du\, u^2 
\ln\left[1-\frac{(1-u\lambda_1/L)(1-u\lambda_2/L)}
{(1+u\lambda_1/L)(1+u\lambda_2/L)}e^{-2u} \right] \, ,
\end{equation}
and the force per area is
\begin{equation}
\label{eq:Force-Robin-pp}
\cF_\|(L) = \frac{\hbar c}{L^4} \Phi'(\lambda_1/L,\lambda_2/L), \quad
\Phi'(\lambda_1/L,\lambda_2/L)= \frac{1}{2\pi^2}\int_0^\infty du\, u^3
\left[1-\frac{(1+u\lambda_1/L)(1+u\lambda_2/L)}
{(1-u\lambda_1/L)(1-u\lambda_2/L)}e^{2u}
\right]^{-1}\, .
\end{equation}
The amplitudes $\Phi(\lambda_1/L,\lambda_2/L)$,
$\Phi'(\lambda_1/L,\lambda_2/L)$ are finite for all
$\lambda_\alpha/L\ge 0$. In general, the integrals in
{Eqs.~\eqref{eq:Energy-Robin-pp}, \eqref{eq:Force-Robin-pp} have}
to be computed numerically.  The PFA predictions as functions of
$\lambda_{1,2}$ have considerable structure, which we illustrate in
Fig.~\ref{fig:plates-amplitude}. 

The behavior of the PFA at asymptotically small or large
$\lambda_{\a}/L$ determines the Casimir interaction as $L\to 0$.
For all non-zero values of $\lambda_{1,2}$,  we take
$\lambda_{\a}/L \to \infty$, but for the Dirichlet case,
$\lambda=0$, the limit $\lambda_{\a}/L\to 0$ applies. For
parallel plates with Robin boundary conditions{,} in the limit
$\lambda_{1,2}/L\gg 1$ we obtain the result for Neumann boundary
conditions on both plates,
\begin{equation}
  \label{eq:pp-amp-rep}
  \Phi(\lambda_1/L,\lambda_2/L) \to \Phi_0^- = -\frac{\pi^2}{1440} \, ,
\end{equation} 
and for $\lambda_{1,2}/L\ll 1$ we obtain the identical result for plates
with Dirichlet conditions.  Finally for 
$\lambda_{1,2}/L\ll 1\ll
\lambda_{2,1}/L$, we obtain the parallel plate result for unlike
(Dirichlet/Neumann) boundary conditions,
\begin{equation}
  \label{eq:pp-amp-att}
  \Phi(\lambda_1/L,\lambda_2/L) \to \Phi_0^+ = -\frac{7}{8} \Phi_0^- =
\frac{7\pi^2}{11520} \, .
\end{equation}
The last case is relevant at short distances if one of the
$\lambda_\alpha=0$.  For two spheres of radius $R$ and
center-to-center separation $d$ with Robin boundary conditions, the
PFA results can now be obtained easily from Eq.~(\ref{eq:energy-pfa}). 
In terms of the surface-to-surface distance $L=d-2R$, we get
\begin{equation}
\label{eq:pfa-spheres}
\cE_\text{PFA} =  \Phi_0^\pm \, \frac{\pi}{2}
\frac{R \, \hbar c}{(d-2R)^2} \, ,
\end{equation}
where the $+$ applies if one and only one $\lambda_\alpha=0$, and the
$-$ in all other cases. Hence, at small separation the interaction
becomes independent of $\lambda_\alpha$, in the sense that it only
depends on whether one $\lambda_\alpha$ is zero. In fact, the form of the
$\mathbb{T}$-matrix of Eq.~\eqref{eq:spheres-t-matrix-imag} shows that
the energy becomes independent of $\lambda_\alpha$ for $d \ll
\min(\lambda_1,\lambda_2)$.  Indeed, if we write the frequency as
$\kappa=u/d$, the translation matrix depends only on $u$, and the
$\mathbb{T}$-matrix can be written as
\begin{equation}
\label{eq:spheres-t-matrix-resc}
\cT^\alpha_{lmlm}=(-1)^l \frac{\pi}{2}
\frac{(d/\lambda_\alpha+d/2R_\alpha) I_{l+1/2}(u R_\alpha/d)
-u  I'_{l+1/2}(u R_\alpha/d)}{(d/\lambda_\alpha+d/2R_\alpha)
K_{l+1/2}(u R_\alpha/d)
-u K'_{l+1/2}(u R_\alpha/d)} \, ,
\end{equation}       
so that the energy depends only on $d/R_\alpha$ for $d\ll
\lambda_\alpha$. This universal behavior is confirmed by a numerical
evaluation of the energy that is described below.

With the results obtained above, we can analyze the sign of the
interaction between plates and spheres at both asymptotically large
and small distances.  Since the PFA result is expected to hold in the
limit where the distance tends to zero, Eq.~(\ref{eq:pfa-spheres})
predicts the sign of the interaction between spheres in the limit of
vanishing distance.  In the limit of large distances, we can compare
the results for parallel planes from Eqs.~(\ref{eq:pp-amp-rep}) and
(\ref{eq:pp-amp-att}) to our calculations for two spheres.  We find
that the {\it sign} of the asymptotic interaction depends on the
choice for $\lambda_\alpha$ and is {\it identical} for plates and
spheres.  Hence, we obtain a complete characterization of the sign of
the interaction at asymptotically large and small distances for the
plate and sphere geometry, which is summarized in
Table~\ref{tab:sign}. However, as we have seen above, the power law
decay at large distance is quite different for plates and spheres.

\begin{table}[ht]
\begin{tabular}{|c|c|c|c|l|}
\hline
$\lambda_1$ & $\lambda_2$ & $L \to 0$ & $L \to\infty$ & remark \\  
\hline
0 & 0 & $-$ & $-$ & $-$ for all $L$ \\
$\infty$ & 0 & $+$ & $+$ & $+$ for all $L$ \\
$\infty$ & $\infty$ & $-$ & $-$ & $-$ for all $L$ \\
$]0,\infty[$ & $]0,\infty[$ & $-$ & $-$ & $+$ at intermediate $L$ for \\
& & & & large enough ratio of $\lambda_1$, $\lambda_2$.\\
& & & & (for plates: $\lambda_1/\lambda_2$ or
$\lambda_2/\lambda_1\gtrsim 2.8$)\\
$]0,\infty[$ & 0 & $+$ & $-$ & \\
$]0,\infty[$ & $\infty$ & $-$ & $+$ & \\
\hline
\end{tabular}
\caption{The sign of the Casimir force between two plates and two
spheres with Robin boundary conditions at asymptotically small and large
surface-to-surface distance $L$. The sign in these two limits is
identical for plates and spheres.  Here ``$-$'' and ``$+$'' indicate
attractive and repulsive forces, respectively.}
\label{tab:sign}
\end{table}

\subsubsection{Casimir forces for all separations}

To go beyond the analytic large distance expansion, we compute
numerically the interaction between two spheres of the same radius $R$
with Robin boundary conditions. Guided by the classification of the
Casimir force according to its sign at small, intermediate and large
separations, we discuss the six different cases listed in
Table~\ref{tab:sign}. Our numerical approach starts from
Eq.~\eqref{finalform2real}.  Using the matrix elements
of Eq.~(\ref{eq:spheres-t-matrix-imag}) and
Eq.~(\ref{eq:spheres-transl-matrix-imag}){,} we compute the
determinant and the integral over imaginary frequency numerically.  We
truncate the matrices at a finite multipole order $l$ so that they
have dimension $(1+l)^2 \times (1+l)^2$, yielding a series of
estimates $\cE^{(l)}$ for the Casimir energy.

$\cE^{(1)}$ gives the exact
result for asymptotically large separations, while for decreasing
separations an increasing number of multipoles has to be included.
The exact Casimir energy at all separations is obtained by
extrapolating the series $\{\cE^{(l)}\}$ to $l\to\infty$. We observe an
exponentially fast convergence as $|\cE^{(l)} - \cE|\sim
e^{-\delta(d/R-2)l}$, where $\delta$ is a constant of order
unity. Hence, as the surfaces approach each other for $d\to 2R$, the
rate of convergence tends to zero. However, we find that the
first $l=20$ elements of the series are sufficient to obtain accurate
results for the energy at a separation with $R/d=0.48$, corresponding
to a surface-to-surface distance of the spheres of $L=0.083 R$, i.e.,
approximately $4\%$ of the sphere diameter. In principle our approach
can be extended to even smaller separations by including higher order
multipoles. However, at such small separations semiclassical
approximations like the PFA start to become accurate and can be also
used.

\begin{figure}[ht]
\includegraphics[scale=0.35]{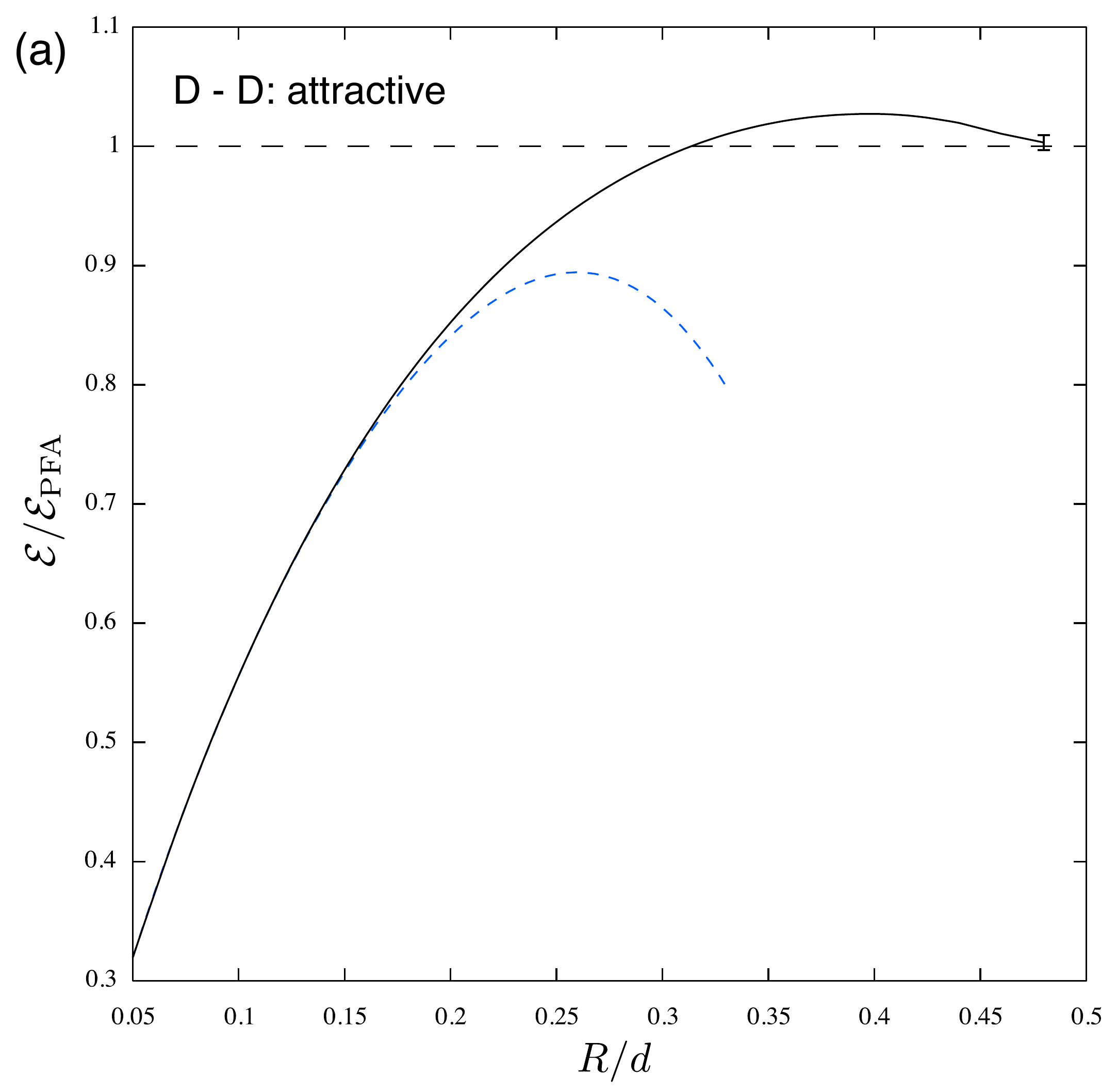}
\includegraphics[scale=0.35]{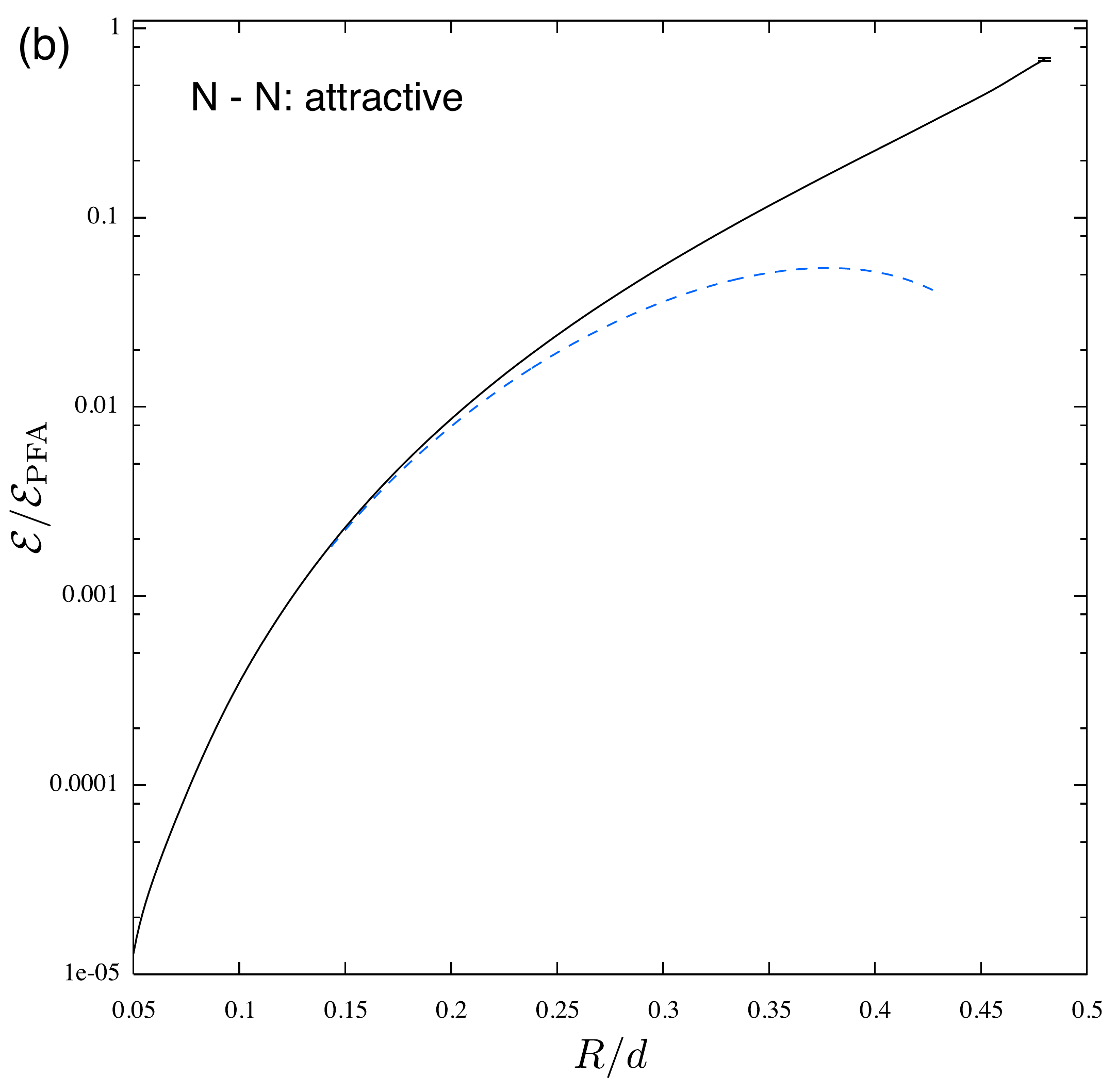}
\includegraphics[scale=0.35]{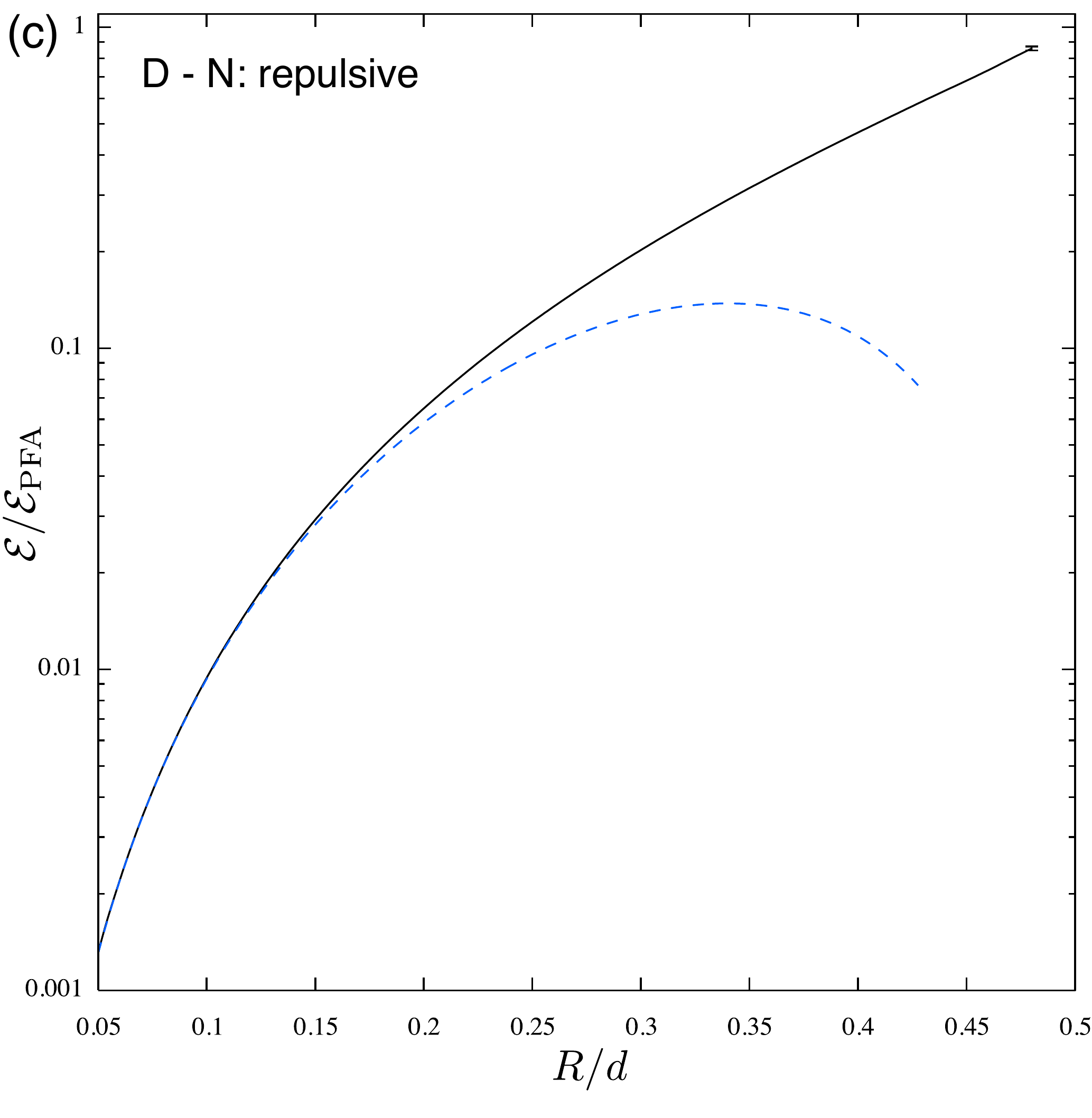}
\caption{(Color online) Casimir energy for two spheres of radius $R$ and
  center-to-center distance $d$: (a) Dirichlet boundary conditions for
  both spheres, (b) Neumann boundary conditions for both spheres, (c)
  Spheres with different boundary conditions (one Dirichlet, one
  Neumann).  The energy is scaled by the PFA estimate of
  Eq.~(\ref{eq:pfa-spheres}). The solid curves are obtained by
  extrapolation to $l\to\infty$. For the smallest separation, the
  extrapolation uncertainty is maximal and indicated by an error
  bar. The dashed curves represent the asymptotic large distance
  expansion given in Eq.~(\ref{eq:spheres-energy-large-d}) with the
  coefficients of Eqs.~(\ref{eq:coeff_D-D}), (\ref{eq:coeff_N-N}) and
  (\ref{eq:coeff_D-N}), respectively.}
\label{fig:numerics-D+N-cases}
\end{figure}

The results for Dirichlet and Neumann boundary conditions are shown in
Fig.~\ref{fig:numerics-D+N-cases}. All energies are divided by
$\cE_\text{PFA}$, given in Eq.~\eqref{eq:pfa-spheres}, with the
corresponding amplitude $\Phi_0^+$ (repulsive at small separations) or
$\Phi_0^-$ (attractive at small separations).  For like boundary
conditions, either Dirichlet or Neumann, the interaction is attractive
at all separations, but  for unlike boundary
conditions it is repulsive.  At large separations the numerical
results show excellent
agreement with the asymptotic expansion derived above.  Note that the
reduction of the energy compared to the PFA estimate at large distances 
depends strongly on the boundary conditions, showing the
different power laws at asymptotically large separations. In the limit
of a vanishing surface-to-surface distance ($R/d\to 1/2$), the energy
approaches the PFA estimate in all cases. Generically, the PFA
overestimates the energy:  $\cE_\text{PFA}$ is approached from
below for $R/d\to 1/2$, except in the case of  Dirichlet boundary
conditions on both spheres, where the PFA underestimates the
actual energy in a range of $0.3 \lesssim R/d < 1/2$. The deviations
from the PFA are most pronounced for Neumann boundary conditions. At a
surface-to-surface distance of $L=3R$ ($R/d=0.2$){, the} PFA
overestimates the energy by a factor of $100$.

\begin{figure}[ht]
\includegraphics[scale=0.17]{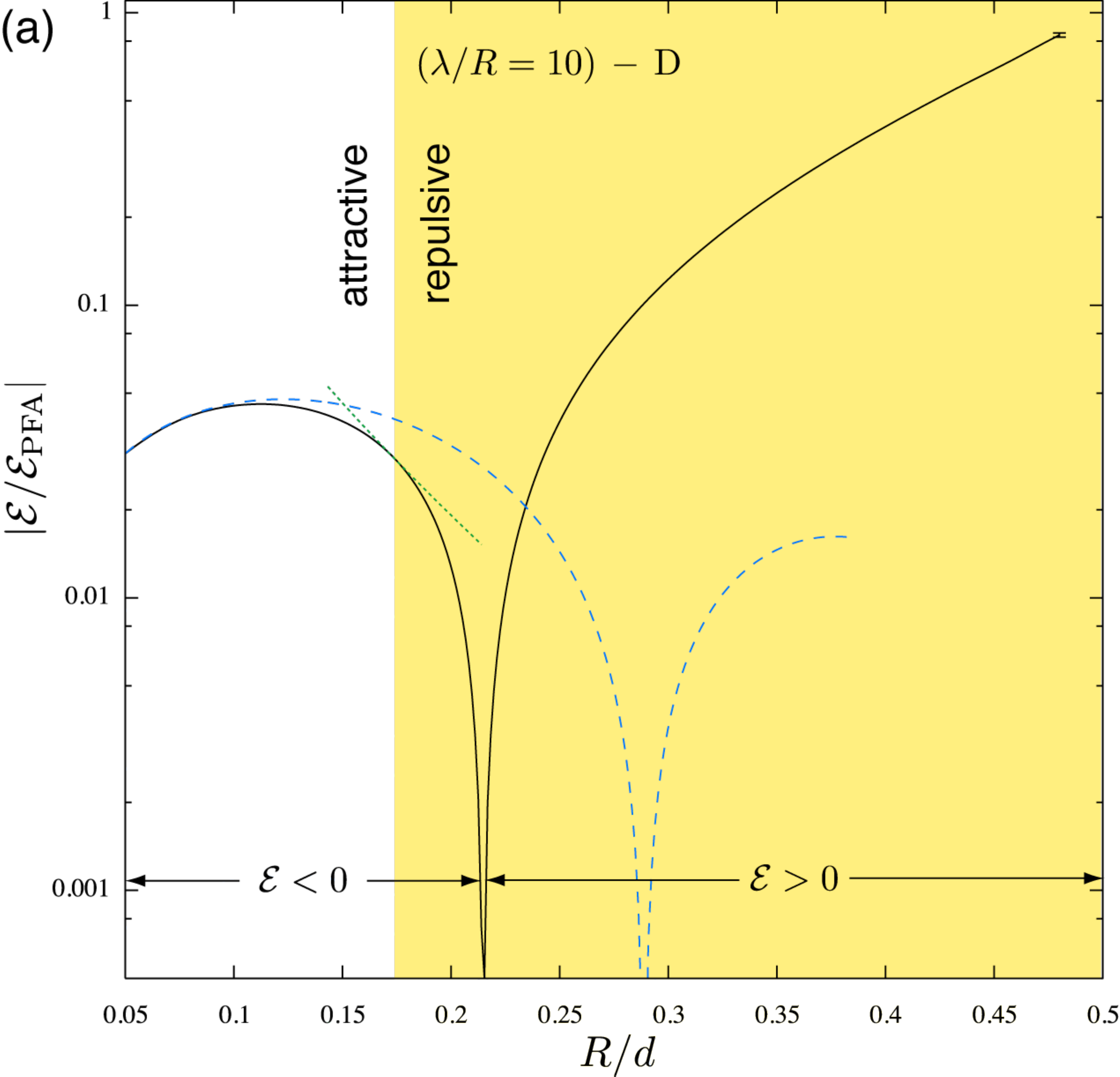}
\includegraphics[scale=0.35]{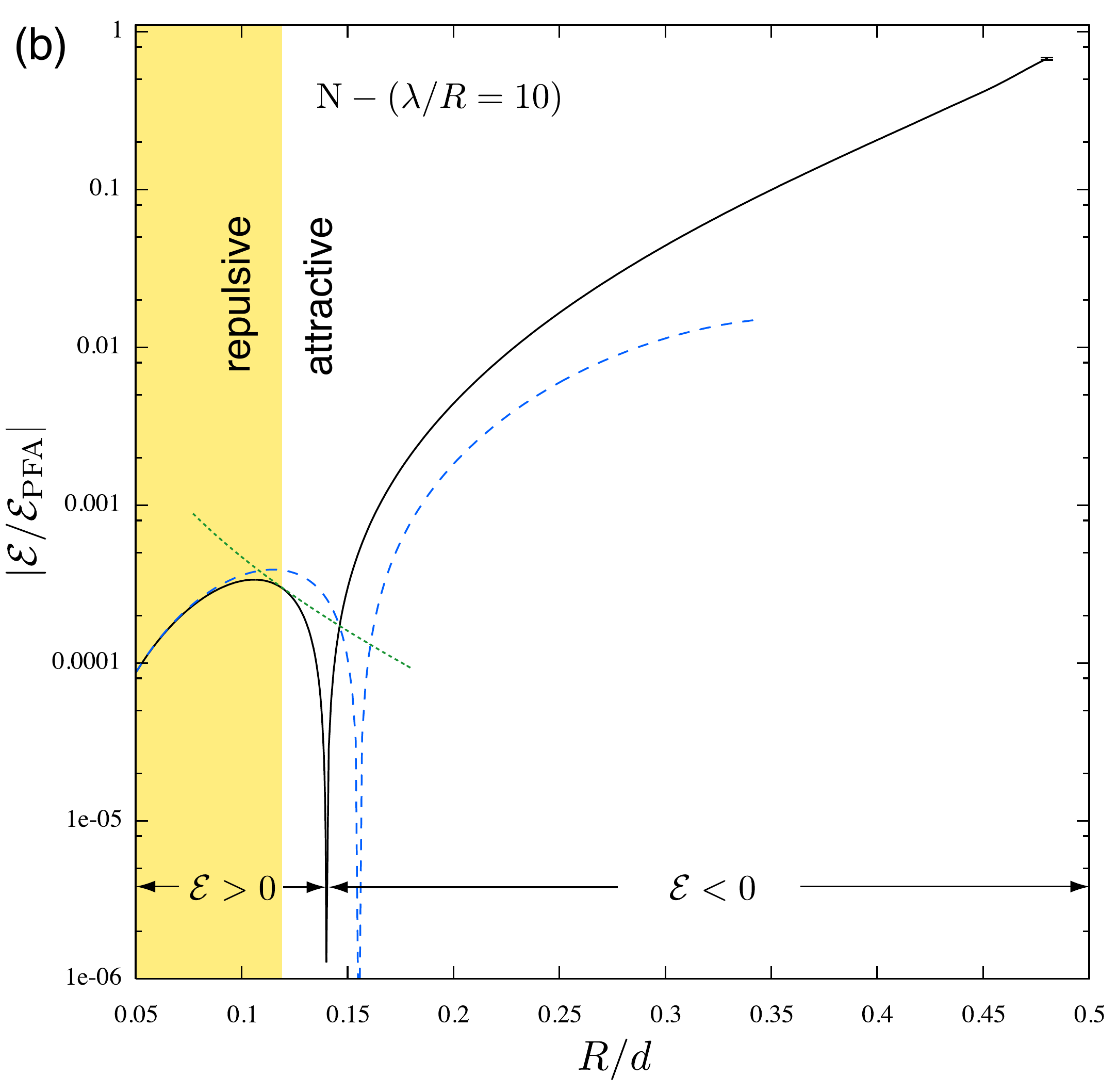}
\includegraphics[scale=0.17]{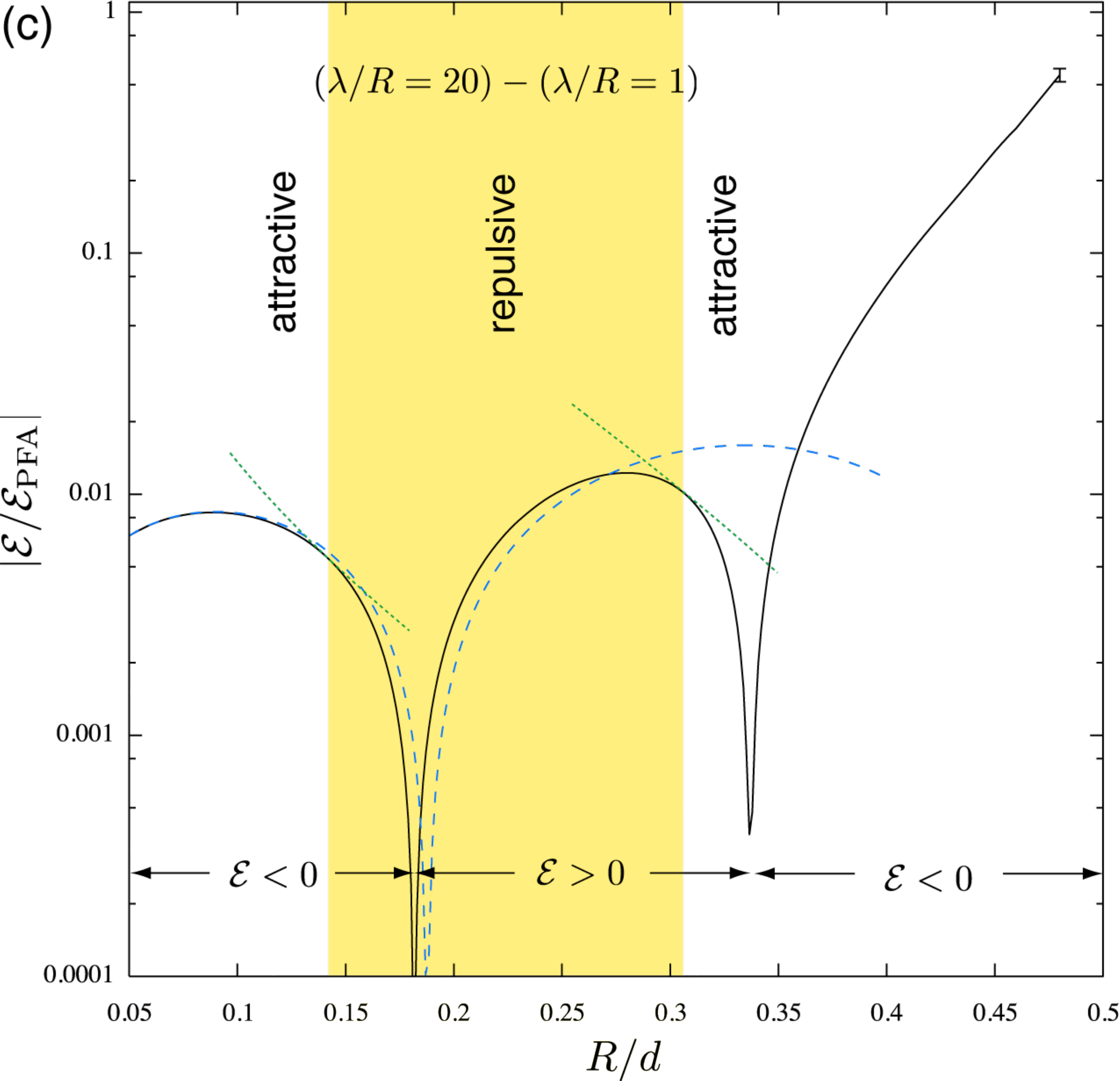}
\caption{(Color online) The Casimir energy for two spheres with different Robin
  boundary conditions for finite $\lambda_\alpha$: (a) Dirichlet
  boundary conditions and $\lambda/R=10$, (b) Neumann boundary
  conditions and $\lambda/R=10$, (c) $\lambda_1/R=10$ and
  $\lambda_2/R=1$.  The solid curves correspond to extrapolated
  results for $l\to\infty$, and the dashed curves represent the
  asymptotic large distance expansion given in
  Eq.~(\ref{eq:spheres-energy-large-d}) with the coefficients of
  Eqs.~(\ref{eq:coeff_D-finite}), (\ref{eq:coeff_N-finite}),
  (\ref{eq:sphere-energy-coeff}), respectively. For logarithmic
  plotting, the modulus of the energy is shown, and the sign of the
  energy is indicated at the bottom. The range of separations with a
  repulsive force is shaded. The points of vanishing force occur where
  an auxiliary function (dotted curves) is tangent to the solid curve,
  see text for details.}
\label{fig:numerics-finite-lambda}
\end{figure}

Casimir interactions for Robin boundary conditions with finite
$\lambda_\alpha$ are shown in
Fig.~\ref{fig:numerics-finite-lambda}. If $\lambda_1=\lambda_2$ the
interaction is always attractive. If the $\lambda_\alpha$ are not
equal and their ratio is sufficiently large, the Casimir force changes
sign either once or twice.  This behavior resembles the interaction of
two plates with Robin boundary conditions.  However, the criterion for
the existence of sign changes in the force now depends not only on
$\lambda_1/\lambda_2$, but on both quantities $\lambda_1/R$ and
$\lambda_2/R$ separately.  Even with $\lambda_1/\lambda_2$ fixed, for
smaller $\lambda_\alpha/R$ there can be sign changes in the force,
while for larger $\lambda_\alpha/R$ the force is attractive at all
distances. When the ratio $\lambda_1/\lambda_2$ is sufficiently large
(or formally infinite for Dirichlet or Neumann boundary conditions),
we can identify three different generic cases where sign changes in
the force occur:
\begin{itemize}
\item 
First, we consider Dirichlet boundary conditions ($\lambda_1=0$)
on one sphere and a finite non-vanishing $\lambda_2/R$ at the other
sphere. Figure~\ref{fig:numerics-finite-lambda}(a) displays the
energy for $\lambda_2/R=10$ as a typical example. At large distances
the energy is negative, while it is positive at short separations
with one sign change in between. The asymptotic expansion of
Eq.~\eqref{eq:spheres-energy-large-d} with the coefficients of
Eq.~\eqref{eq:coeff_D-finite} yields the exact energy at separations
well below the sign change. While the expansion predicts
qualitatively the correct overall behavior of the energy, it does
not yield the actual position of the sign change correctly. Of
course, for the Casimir interaction between compact objects, the
sign of the force $\cF=-\partial \cE/\partial d$ is
the physically important quantity, not the energy.  The distance
at which the force vanishes cannot be deduced directly from the slope
of the curve for $\hat\cE\equiv \cE/\cE_\text{PFA}$, since one has
\begin{equation}
  \label{eq:force-from-energy-ratio}
  \hat\cE'(d) = \frac{1}{\cE_\text{PFA}(d)} \left[
\cE'(d) +\frac{2}{d-2R} \cE(d)
\right] \, .
\end{equation}
The force vanishes at the distance $d_0$ if $\cE'(d_0)=0$, so that
\begin{equation}
\label{eq:zero-force-cond}
\hat\cE'(d_0)=\frac{2}{d_0-2R} \hat \cE(d_0) \, .
\end{equation}
Hence the distance at which the force vanishes is determined by the
position $d_0$ where the curve of the auxiliary function $t(d)=\tau
(d/R-2)^2$ is tangent to the curve of $\hat \cE$. The two unknown
quantities $d_0$ and $\tau$ are then determined by the conditions
$\hat \cE(d_0)=t(d_0)$ and $\hat \cE'(d_0)=t'(d_0)$.  This procedure
allows us to obtain the distance at which the force vanishes easily{,}
without computing derivatives numerically. The tangent segment of the
curve for $t(d)$ is shown in Fig.~\ref{fig:numerics-finite-lambda}(a)
as a dotted line.  From this construction we find that at a distance
$d_{-\Rightarrow +}$ the force changes from attractive to repulsive
for decreasing separations.  The position $d_{-\Rightarrow +}$
corresponds to a minimum of the energy and decreases with decreasing
$\lambda_2/R$, so that in the limit $\lambda_2/R\to 0$ it approaches
the case of two spheres with Dirichlet boundary conditions, where the
force is always attractive.

\item Second, we study Neumann boundary conditions on one sphere and a
  inite non-vanishing $\lambda_2/R$ at the other sphere.  As an
  example we choose again $\lambda_2/R=10$, as shown in
  Fig.~\ref{fig:numerics-finite-lambda}(b). The energy is positive at
  large distances and becomes negative at small distances.  The
  asymptotic expansion with the coefficients of
  Eq.~\eqref{eq:coeff_N-finite} is found to be valid well below the
  separation where the sign of the energy changes.  Hence, the
  expansion describes the behavior of the energy qualitatively, but
  does not predict the precise position of the sign change. The sign
  change of the force can be obtained by the method described above.
  At a position $d_{+\Rightarrow -}$, the force changes from repulsive
  to attractive with decreasing separation and the energy is maximal.
  A decreasing (increasing) $\lambda_2/R$ shifts $d_{+\Rightarrow -}$
  to smaller (larger) separations.  This result is consistent with an
  entirely repulsive (attractive) force for Neumann-Dirichlet
  (Neumann-Neumann) boundary conditions.

\item
 The third case is obtained if both $\lambda_\alpha$ are finite
  and non-zero. A typical example with $\lambda_1/R=20$ and
  $\lambda_2/R=1$ is shown in
  Fig.~\ref{fig:numerics-finite-lambda}(c). The energy is negative
  both at large and small separations but turns positive at
  intermediate distances.  The asymptotic expansion applies again at
  sufficiently large separations beyond the position where the energy
  becomes positive. For values of the ratio $\lambda_1/\lambda_2$ that
  are larger than an $R$-dependent threshold, the force changes sign
  twice, so that it is repulsive between the separations
  $d_{-\Rightarrow +}$ and $d_{+\Rightarrow -}$.  The energy has a
  minimum (maximum) at $d_{-\Rightarrow +}$ ($d_{+\Rightarrow -}$).
  If $\lambda_1/R$ increases and $\lambda_2/R$ decreases, the
  repulsive region grows until eventually the force becomes repulsive
  at all separations, corresponding to the limit of Dirichlet/Neumann
  boundary conditions.  Decreasing $\lambda_1/R$ and increasing
  $\lambda_2/R$ reduces the interval with repulsion.  In this case,
  first the zeros of the energy disappear, leaving negative energy at
  all distances but still a repulsive region, and then the two
  positions where the force vanishes merge, leaving an entirely
  attractive force.
\end{itemize}

\subsection{Interaction in terms of low frequency scattering data}
\label{sec:inter-terms-low}

We noted in Sec.~\ref{2.3} that in the limit of vanishing frequency,
the $\mathbb{S}$-matrix is related to tensor generalizations of the
capacitance.  At higher frequencies, this tensor receives corrections
that can be expanded in frequency.  The object's shape is thus encoded
in the tensor expansion coefficients of the $\mathbb{S}$-matrix.  We
can use these results to obtain the Casimir interaction between
non-spherical objects in a large distance expansion.  From such
{an} analysis one can obtain information about which pieces of the
$\mathbb{S}$-matrix contribute to the Casimir interaction at a given
order in inverse separation.

In general, according to Eq.~\eqref{eq:scatter}, the $\mathbb{T}$-matrix of
object $\alpha$ can be written in a small $\kappa$-expansion of the
form
\begin{equation}
  \label{eq:T-small-kappa}
  \cT^\alpha_{l'm'lm} = \frac{i^{l+l'}}{(2l+1)!!(2l'-1)!!} \sum_{q=0}^\infty
 C^\alpha_{q;l'm'lm} \kappa^{l+l'+1+q} \, .
\end{equation}
Due to the symmetry of the $\mathbb{T}$-matrix, the coefficients obey the relation
\begin{equation}
  \label{eq:C-relation}
  C^\alpha_{q;l'm'lm} = \frac{2l+1}{2l'+1}  C^\alpha_{q;lml'm'} \, .
\end{equation}
We assume again that the origins of the two objects are located on the
$z$-axis at a separation $d$. Then we can use the translation matrix
elements of Eq.~\eqref{eq:spheres-transl-matrix-imag}. The Casimir
energy can be obtained from the low-$\kappa$ expression for the
$\mathbb{T}$-matrix of Eq.~\eqref{eq:T-small-kappa} by a simultaneous
expansion in the number of partial waves, {the} number of
scatterings (powers of ${\mathbb N}$), and {the} imaginary
frequency $\kappa$. The analysis based on the general low-frequency
structure of the $\mathbb{T}$-matrix for objects of arbitrary shape
and boundary condition becomes increasingly
complicated with decreasing separation between the
objects. Since the $\mathbb{T}$-matrix cannot be assumed to be
diagonal in $l$, partial waves of order $l$ can contribute to the
energy at order $d^{-(3+l)}$ (for $\mathbb{T}$-matrices
that are diagonal in $l$, they can only contribute at order
$d^{-(3+2l)}$).  We consider here only partial waves up to order $l=3$
and four scatterings, which allows us to obtain the energy to order
$d^{-6}$. The energy can again be written in the form
\begin{equation}
  \label{eq:energy_general_shape}
  \cE = \frac{\hbar c}{\pi} \sum_{j=3}^\infty \frac{B_j}{d^j} \, .
\end{equation}
The coefficients $B_j$ are completely specified by the low-frequency
behavior of the $\mathbb{T}$-matrix, i.e., they can be obtained from the
coefficients $C^\alpha_{q;l'm'lm}$ in Eq.~\eqref{eq:T-small-kappa}. In
the following, we assume that the origin of the object's coordinate
frame has been chosen such that the dipole response to an applied
constant potential vanishes, i.e.,
$C^\alpha_{0;1m00}=C^\alpha_{0;001m}=0$ for $m=-1,0,1$.  Using the
symmetry relation of Eq.~\eqref{eq:C-relation}, the result for the
coefficients can be written as
\begin{eqnarray}
\label{eq:general-shape-coeff}
 B_3 &=& -\frac{1}{4}C^1_{0;0000}C^2_{0;0000} ,\\
 B_4 &=& 
-\frac{1}{8} \left( C^1_{0;0000}C^2_{1;0000} + C^2_{0;0000}C^1_{1;0000}
\right) ,\\
 B_5 &=& -\frac{1}{16} [C^1_{0;0000}C^2_{0;0000}]^2 \nonumber\\
&&-\frac{5}{8} \left(C^1_{0;0000} C^2_{0;1010} + C^2_{0;0000}
C^1_{0;1010} \right)
-\frac{\sqrt{5}}{6}\left(C^1_{0;0020}C^2_{0;0000}+C^2_{0;0020}C^1_{0;0000}
\right)\nonumber\\
&&-\frac{1}{2\sqrt{3}} \left(C^1_{0;0000}C^2_{1;0010}-C^2_{0;0000}C^1_{1;0010}
\right)
-\frac{1}{8} \left(C^1_{1;0000}C^2_{1;0000} +C^1_{0;0000}C^2_{2;0000}
+ C^2_{0;0000}C^1_{2;0000} \right) ,
\\
B_6 &=& 
-\frac{1}{32} \left( C^1_{0;0000}C^1_{1;0000} [C^2_{0;0000}]^2 +
C^2_{0;0000}C^2_{1;0000} [C^1_{0;0000}]^2
\right)\nonumber\\
&& -\frac{7}{8}\sqrt{\frac{5}{3}}\left(C^1_{0;0000}C^2_{0;1020}
-C^2_{0;0000}C^1_{0;1020}
\right)
-\frac{\sqrt{7}}{8}\left(C^1_{0;0000}C^2_{0;0030} - C^2_{0;0000}C^1_{0;0030} 
\right)\nonumber\\
&& -\frac{9}{16}\left(C^1_{0;0000}C^2_{1;1010} +
C^2_{0;0000}C^1_{1;1010} + C^1_{1;0000}C^2_{0;1010}
+ C^2_{1;0000}C^1_{0;1010} 
\right)
\nonumber\\
&& -\frac{\sqrt{5}}{8}\left(C^1_{1;0020} C^2_{0;0000} + C^2_{1;0020}
C^1_{0;0000} + C^1_{1;0000} C^2_{0;0020}+C^2_{1;0000} C^1_{0;0020}  
\right)
\nonumber\\
&& -\frac{5}{8\sqrt{3}}\left(
C^1_{0;0000}C^2_{2;0010} - C^2_{0;0000}C^1_{2;0010} 
+C^1_{1;0000}C^2_{1;0010}-C^2_{1;0000}C^1_{1;0010}   
\right)\nonumber\\
&& -\frac{3}{16} \left(
C^1_{1;0000}C^2_{2;0000} + C^2_{1;0000}C^1_{2;0000} +
C^1_{0;0000}C^2_{3;0000} + C^2_{0;0000}C^1_{3;0000}
\right) \, .
\end{eqnarray}
This result deserves several comments. First, the above expressions
for the coefficients have been substantially simplified by the
requirement that $C^\alpha_{1m00}=C^\alpha_{001m}=0$, which eliminates
$2$, $5$ and $22$ terms from $B_4$, $B_5$ and $B_6$, respectively.
Most contributions originate from two scatterings, which can be seen
from the appearance of a product of two coefficients
$C^\alpha_{q;lml'm'}$.  Only the leading terms of $B_5$ and $B_6$ are
composed of a product of four coefficients $C^\alpha_{q;lml'm'}$ and
hence result from four scatterings.  The latter terms are completely
determined by the capacitance of the objects and its finite frequency
corrections, $C^\alpha_{q;0000}$. The result clearly shows the
relevant number of partial waves, which increases by one with each
additional power of $1/d$. At leading order, $\sim 1/d^3$, the
coefficient $B_3$ is given by the product of the capacitances
$C^\alpha_{0;0000}$ of the objects. At the next order, contributions
from $l=1$ partial waves are absent due to the choice of origin.
At higher orders, the coefficients $B_j$ contain contributions from
partial waves up to $l=j-3$. Note also that not all terms are
symmetric under an exchange of the two objects,
$C^1_{q,lml'm'}\leftrightarrow C^2_{q,lml'm'}$. The terms for which
$l+l'$ is an odd number change sign under an exchange of the
objects. This is related to the fact that the expansion of the
$\mathbb{T}$-matrix in Eq.~\eqref{eq:T-small-kappa} assumes that the local
coordinate systems of the two objects have the same orientation.
Hence, the objects ``see'' each other along different directions,
i.e., along the direction of the positive and negative $z$-axis,
respectively. Up to coefficient $B_6$, all
terms have $m=m'=0$, which is again due to the choice of origins
eliminating all terms with nonzero $m$ up to the order considered
here. 

The above result can be applied to two objects of arbitrary shape and
orientation{,} since the direction connecting the origins of the objects
can be always defined as the $z$-axis. The orientation of the objects
with respect to this axis is then specified by the coefficients
$C^\alpha_{q,lml'm'}${,} which depend on the angle that a reference
direction, fixed to the object, forms with the $z$-axis.  Hence, the
dependence of the Casimir interaction on the orientation of the
objects can be obtained from this general result by computing the
low-frequency expansion of the $\mathbb{T}$-matrix elements for objects with an
arbitrary orientation with respect to the $z$-axis. Since the
capacitance coefficients $C^\alpha_{q;0000}$ are independent of
orientation, a dependence on orientation can occur only at order
$d^{-5}$. A more detailed study of the orientation dependence and
the reduction of the number of independent expansion coefficients due
to symmetries of the objects is left to a future publication.

\section{Acknowledgments}
This work was supported by the National Science Foundation (NSF)
through grants DMR-04-26677 (MK) and PHY-0555338 (NG), by a Cottrell
College Science Award from Research Corporation (NG), by the
U.~S.~Department of Energy (DOE) under cooperative research agreement
\#DF-FC02-94ER40818 (RLJ), and by a Heisenberg Fellowship from the
German Research Foundation (TE).

\begin{appendix}

\section{Neumann Boundary Conditions}
\label{app:B}

Throughout the paper we considered Dirichlet boundary conditions,
$\phi=0$.  In fact, our final result, Eq.~\eqref{finalform}, which
expresses the interaction Casimir energy of many compact objects in
terms of the transition and translation matrices, is independent of
the choice of boundary conditions, provided the $\mathbb{T}$-matrix is
the one appropriate to the boundary conditions of interest.  Here we
show this {result} for the Neumann case.  The general Robin case
is left to the reader.

Neumann boundary conditions are implemented by replacing $\phi(\bfx )$
by $\p_{n}\phi(\bfx )$ in Eq.~\eqref{deltafn}, leading to an
expression for the partition function analogous to Eq.~\eqref{source3}
with $\phi(\bfx )\to \p_{n}\phi(\bfx )$.  Like the Dirichlet case,
this case also has an analogy to electrostatics, namely
a complex field coupled to a set of \emph{surface dipole densities}
 $\varrho_\alpha(\bfx)$.  By analogy, in this case $\p_{n}\phi$ will be
continuous throughout space, but $\phi$ itself will jump by
$\varrho_{\a}(\bfx )$ across the surface $\Sigma_{\a}$.  Therefore the
classical equations of motion analogous to Eq.~\eqref{variation} are
\begin{eqnarray}
-\left(\nabla^{2} +k^{2}\right)\phicl(\bfx)&=& 0,\quad\mbox{for $\bfx\notin \Sigma_\a$},\nonumber\\
\Delta \phicl(\bfx )&=& -\varrho_{\a}(\bfx ),\quad\mbox{for $\bfx\in \Sigma_\a$},\nonumber\\
\left.\Delta \p_{n}\phicl\right|_{\bfx }&=&0,\quad \mbox{for $\bfx\in \Sigma_\a$} \, ,
\label{b.1}
\end{eqnarray}
  and the normal
derivative of the free Green's function generates the field associated
with these dipole sources,
\begin{equation}
\label{b.2}
\phicl(\bfx) =\sum_{\beta}\int_{\Sigma_{\b}}d\bfx'\p_{n'}
\cG_{0}(\bfx,\bfx',k)\varrho_{\b}(\bfx') \, ,
\end{equation}
where $\p_{n'}$ denotes the normal derivative at the point $\bfx'$ and acts only on $\cG_{0}$.

The evaluation of the classical action proceeds in analogy to the Dirichlet case,
\begin{equation}
\widetilde S_{\rm cl}[\varrho]=\frac{1}{2}\sum_{\a,\b}\left(\int_{\Sigma_\a} d\bfx \varrho^{*}_{\a}(\bfx )\p_{n}\phi_{\beta}(\bfx )+\cc\right) \, .
\label{b.3}
\end{equation}
First we evaluate the terms with $\a\ne \b$,
\begin{equation}
\label{b.31}
\widetilde S_{\b\a}=\frac{1}{2}\int_{\Sigma_\a} d\bfx \left( \varrho_{\a}^{*}(\bfx )
\p_{n}\phi_{\beta}(\bfx)+\cc\right) \, .
\end{equation}
In analogy to Eq.~\eqref{phib2},
\begin{equation}
\label{b.4}
\phi_{\b}(\bfx)= {ik}\sum_{lm} h^{(1)}_{l}(k r_{\b})Y_{lm}(\hat\bfx_{\b})\int_{\Sigma_{\b}}d\bfx_{\b}'\p_{{n'}}[j_{l}(k r_{\b}')
Y^{*}_{lm}(\hat\bfx_{\b}')]\varrho_{\b}(\bfx_{\b}') \, ,
\end{equation}
which serves to define  multipole moments of the dipole layer density,
\begin{equation}
\label{b.5}
P_{\b,lm}\equiv
\int_{\Sigma_{\b}}d\bfx_{\b}\p_{n}\left[j_{l}(k r_{\b})
Y^{*}_{lm}(\hat\bfx_{\b})\right]\varrho_{\b}(\bfx_{\b})\, . 
\end{equation} 
As in the Dirichlet case, the Hankel functions defined with respect to
$\cO_{\b}$ can be expanded in terms of Bessel functions about
$\cO_{\a}$ with the help of translation formulas.  Substituting into
\Eq{b.31} leads to a result analogous to \Eq{offdiag},
\begin{equation}
\label{b.6}
\widetilde S_{\b\a}[P_{\a},P_{\b}]=
{\frac{ik}{2}}\sum_{lml'm'} P^{*}_{\a,l'm'}\cU^{\alpha\beta}_{l'm'lm}
P_{\b,lm}+\cc  \, .
\end{equation}

When $\alpha=\beta$, we also proceed in analogy to the Dirichlet case.
The interior field is defined as in \Eq{self2} and substitution into
\Eq{b.3} leads to
\begin{equation}
\label{b.7}
\widetilde S_{\a}[\varrho_{\a}]=\frac{1}{2}
\sum_{lm}\left( \phi_{\a,lm}P^{*}_{\a,lm}+\cc \right)\, ,
\end{equation}
in analogy to \Eq{self3}. As in the Dirichlet case, we determine the
field expansion coefficients, $\phi_{\a,lm}$, by constructing the
exterior field $\phi_{{\rm out},\a}$ in two different ways.  First we
construct it following the logic of Eqs.~(\ref{outsidereg})--(\ref{outagain}), 
but now with the aid of the \emph{Neumann}
$\mathbb{T}$-matrix , since the normal derivative of $\phi_{\a}$ must be
continuous across $\Sigma_{\a}$.  Then we construct 
$\phi_{{\rm out},\a}$ directly from the Neumann multipole moments using
\Eq{b.2}.  Equating the two we obtain
 \begin{equation}
\phi_{\a,lm}=-ik\sum_{l'm'}[\cT^{\a}]^{-1}_{lml'm'}P_{\a,l'm'},
\end{equation}
where $\cal T^\alpha$ is the Neumann $\mathbb{T}$-matrix.  Substituting
into \Eq{b.7} we obtain the final result,
\begin{equation}
\label{b.8}
\widetilde S_{\a}[P_{\a}]={-\frac{ik}{2}}
\sum_{lm} P^{ *}_{\a,lm}[\cT^{\a}]^{-1}_{lml'm'}P_{\a,l'm'} + \cc ,
\end{equation}
analogous to \Eq{selfaction}.   The
rest of the derivation follows the Dirichlet case closely.

\section{Casimir energy and functional integrals} 
\label{app:A}
 
Here we show the equivalence between our functional integral formalism
and the standard representation of the Casimir energy in terms of
zero-point energies.  We consider a real scalar field,
$\varphi(\bfx,t)$, that obeys the wave equation in a domain $\cD$, and
obeys a Dirichlet boundary condition on $\Sigma$, the boundary of
$\cD$, which need not be connected.  Let $\phi_{\a}(\bfx)$ and
$-\varepsilon_{\alpha}^{2}/\hbar^{2}c^{2}$ be the eigenfunctions and
eigenvalues of the Laplacian in $\cD$,
\begin{eqnarray}
\label{helm}
-\nabla^{2}\phi_{\alpha}(\bfx)&=&\frac{\varepsilon_{\alpha}^{2}}{\hbar^{2}c^{2}}\phi_{\alpha}(\bfx)\quad\mbox{in}\quad \cD,\nonumber\\
\phi_{\alpha}(\bfx)&=&0\quad\mbox{on}\quad \Sigma \, .
\end{eqnarray}
Let $-\varepsilon^{2}_{\infty,\alpha}/\hbar^{2}c^{2}$ be the
corresponding eigenvalues when the objects are moved to arbitrarily
large separation from one another. For simplicity we assume that the
spectrum is discrete in both cases.
 
Consider the Minkowski space functional integral over the field
$\varphi(\bfx,t)$, as introduced in \Eq{effact} but now for a
real field,
\begin{equation}
\label{paths}
{Z}[\cC]= \int \left[\cD\varphi\right]_{\cC}\exp\left\{
\frac{i}{\hbar}\int_0^{T}dt\int_{\cD} d\bfx 
\left(\frac{1}{c^{2}}(
\partial_t\varphi)^{2} -
(\nabla\varphi)^{2}
\right)\right\}\, .
\end{equation}
By definition, only periodic paths, $\varphi(\bfx,0)=\varphi(\bfx,T)$,
that obey the Dirichlet boundary conditions are included.  Since
$\varphi(\bfx,t)$ is periodic in $T$, it can be expanded in a Fourier
series,
$$
\varphi(\bfx,t)=\sum_{n=-\infty}^{\infty} \varphi_{n}(\bfx)
e^{-2\pi nit/T}  \, ,
$$
with $\varphi_{-n}(\bfx) = \varphi_{n}^\ast(\bfx)$.
Because the boundary conditions are time independent, the functional
integral over each frequency mode of $\varphi$ can be done
independently.  Therefore $Z[\cC]$ can be written as an infinite
product,
\begin{equation}
\label{product}
Z[\cC]
=\prod_{n=-\infty}^{\infty}\int
\left[\cD\varphi_{n} \right]_{\cC}\exp \left\{
\frac{iT}{\hbar} \int_{\cD} d\bfx \left(
\left(\frac{2 \pi n}{cT}\right)^{2}
\left|\varphi_{n}\right|^{2}-\left|\nabla\varphi_{n}\right|^{2}\right)\right\} \, .
\end{equation}
At the end of this calculation we will divide by $Z_{\infty}$, the
functional integral defined when all the objects are removed to
arbitrarily large separation.  Therefore we ignore all multiplicative
factors in $Z[\cC]$ that are independent of the location of the
constraining surfaces, beginning with the Jacobian
generated by the change of variables from $\varphi(\bfx,t)$ to
$\varphi_{n}(\bfx)$.
 
The real eigenfunctions, $\{\phi_{\alpha}(\bfx)\}$, defined in
Eq.~\eqref{helm} form a complete orthonormal set of functions on
$\cD$, so the field $\varphi_{n}(\bfx)$ can be expanded as
$$
\varphi_{n}(\bfx)=\sum_{\alpha}c_{n\alpha}\phi_{\alpha}(\bfx) \, ,
$$
where $\{c_{n\alpha}\}$ are complex numbers 
with $c_{-n\alpha}= c_{n\alpha}^\ast$\,.  When this
expansion is substituted into $Z[\cC]$,
the functional integral reduces to an infinite product of ordinary
integrals over the coefficients $\{c_{n\alpha}\}$,
\begin{equation}
\label{in}
Z[\cC] = \prod_{n=-\infty}^{\infty}
\prod_{\alpha}\int dc_{n\alpha} 
\exp\left\{-\frac{iT}{\hbar}\sum_{\alpha}
\left(\frac{\varepsilon_\alpha^{2}}{\hbar^{2}c^{2}}-
\left(\frac{2\pi n}{cT}\right)^{2}\right)
|c_{n\alpha}|^2\right\}\, ,
\end{equation}
where  we have dropped  factors common to $Z[\cC]$ and $Z_{\infty}$.
The integrals over the $\{c_{n\alpha}\}$ can be performed (taking $T$
complex with a negative imaginary part, and analytically continuing
back to $T$ real at the end), leading to
\begin{equation}
Z[\cC] = \prod_{n=-\infty}^{\infty}
\prod_{\alpha} \left(\frac{\varepsilon_\alpha^{2}}{\hbar^{2}c^{2}}-
\left(\frac{2\pi n}{cT}\right)^{2}\right)^{-1/2} \, .
\end{equation}
Then, interchanging the products over $n$ and $\alpha$,
$Z[\cC]$ can be written,
\begin{equation}
Z[\cC] = \prod_{\alpha}
\frac{\hbar c}{\varepsilon_{\alpha}}
\prod_{n=1}^{\infty}
\left[ \frac{\varepsilon_\alpha^{2}}{\hbar^{2}c^{2}}-
\left(\frac{2\pi n}{cT}\right)^{2} 
\right]^{-1} \, ,
\end{equation}
where we have separated out the $n=0$ term and combined each positive
and negative $n$ into a single term.  Factoring out the configuration
independent factor
$\prod_{n=1}^{\infty}\left(\frac{2\pi n}{cT}\right)^{2}$
and using the infinite product representation
of the sine function,
$$
\frac{\sin x}{x}=\prod_{n=1}^{\infty}\left(1-\frac{x^{2}}{\pi^{2}n^{2}}\right)  \, ,
$$
we obtain 
\begin{eqnarray}
\label{almost}
Z[\cC] &=& \prod_{\alpha}\frac{1}{\sin
\frac{\varepsilon_{\alpha}T}{2\hbar}}\nonumber\\
&=&\prod_{\alpha}\
\frac{
e^{-i\frac{\varepsilon_{\alpha}T}{2\hbar}}}{
1-e^{-i\frac{\varepsilon_{\alpha}T}{\hbar}}} \, .
\end{eqnarray}
Finally we define $T=-i\Lambda/c$, send $\Lambda\to\infty$, divide by
$Z_{\infty}$, and take the logarithm, obtaining
\begin{equation}
\lim_{\Lambda \to\infty}\ln\left(Z[\cC]/Z_{\infty}\right) = 
-\frac{\Lambda}{2\hbar c}\sum_{\alpha}(
\epsilon_{\alpha} - \epsilon_{\alpha,\infty}) \, ,
\end{equation}
which, using \Eq{casimir1}, gives the standard expression for the
Casimir energy of a real scalar field,
\begin{equation}
\label{a.1}
\cE[\cC]= \frac{1}{2}\sum_{\alpha}\left(
\varepsilon_{\a}-\varepsilon_{\infty,\a} \right)\, .
\end{equation}
Note that if we had set $T=-i\beta\hbar$ and not taken the
$T\to\infty$ limit, we would have obtained the partition function for
a Bose ``gas'' of scalar particles occupying the eigenstates with
energy $\varepsilon_{\a}$, the starting point for a quantum
statistical mechanical treatment of the Casimir effect at 
temperature $1/\beta$.

\section{Translation Formulas}
\label{app:C}

For completeness, we sketch the derivations of the translation formulas
quoted in Eqs.~\eqref{eq:reg+out+solutions} and \eqref{forms}
{with $\beta=1$ and $\alpha=2$}.  See Fig.~\ref{objects} for the
configuration of the two objects. 

\subsection{$\cV^{12}_{l'm'lm}(\bfX_{12})$}
\label{app:C1}

First consider the equation relating the regular partial wave solution
about $\bfx_{2}$ to those expanded about ${\cal O}_{1}$.  Since
$\bfx_{2}=\bfX_{21}+\bfx_{1}$ we can write 
\begin{equation} 
\label{tr.1}
e^{i\bkk\bfx_{2}}=e^{i\bkk\bfX_{21}}e^{i\bkk\bfx_{1}} ,
\end{equation}
for an arbitrary vector $\bkk$.  If we use the familiar partial wave
expansion, 
\begin{equation} 
\label{tr.2}
e^{i\bkk\bfx}=4\pi\sum_{lm}i^{l}j_{l}(kr)Y^{*}_{lm}(\hat\bkk)
Y_{lm}(\hat\bfx) ,
\end{equation}
in all three terms, multiply by the factor
$Y_{lm} (\bkk)$ and integrate over $\hat \bkk${,} we obtain 
\begin{equation} 
\label{tr.3}
j_{l}(kr_{2})Y_{lm}( \hat\bfx_{2})=4\pi\sum_{l'm'l''m''}i^{l'+l''-l}\left(\int
  d\hat\bkk Y_{lm}(\hat\bkk)Y_{l'm'}^{*}(\hat
  \bkk)Y_{l''m''}^{*}(\hat\bkk)\right)j_{l''}(kd_{12})j_{l'}(kr_{1})Y_{l''m''}
(\hat\bfX_{21})Y_{l'm'}(\hat\bfx_{1}) \, .
\end{equation}
The integral over three spherical harmonics is~\cite{edmonds} 
\begin{equation}
\label{tr.31}
\int d\hat\bkk Y_{lm}(\hat\bkk)Y^{*}_{l'm'}(\hat \bkk)Y^{*}_{l''m''}(\hat\bkk)=(-1)^{m'+m''}
\sqrt{ \frac{\lambda\lambda'\lambda''}{4\pi}}
 \threej{l}{l'}{l''}{0}{0}{0}
\threej{l}{l'}{l''}{m}{-m'}{-m''} ,
\end{equation}
where $\lambda\equiv 2l+1$.  Substituting into \Eq{tr.3} and
regrouping terms we find,
\begin{eqnarray}
\label{tr.3.1}
j_{l}(kr_{2})Y_{lm}( {\hat\bfx_{2}})&=&\sum_{l'm'}\left[\sum_{l''m''}i^{l''}
(-1)^{m''}\sqrt{\lambda''} \threej{l}{l'}{l''}{0}{0}{0}
\threej{l}{l'}{l''}{m}{-m'}{-m''}j_{l''}(kd_{12})Y_{l''m''}
(\hat\bfX_{21})\right] {\times} \nonumber\\
&&
\sqrt{4\pi\lambda\lambda'}i^{l'-l}(-1)^{m'}j_{l'}(kr_{1})Y_{l'm'}({\hat\bfx_{1}}) \, .
\end{eqnarray}
Comparing this {expression} with the definition of the translation matrix,
$\mathbb{V}$, given in \Eq{eq:reg+out+solutions}, we can read off the
matrix elements $\cV^{12}_{lml'm'}(\bfX_{12})$,
\begin{equation}
\label{tr.3.2}
\cV^{12}_{{l'm'lm}}(\bfX_{12})=\sqrt{4\pi\lambda\lambda'}\,
i^{l'-l}(-1)^{m'}\sum_{l''m''}i^{l''}(-1)^{m''}\sqrt{\lambda''}
\threej{l}{l'}{l''}{0}{0}{0}
\threej{l}{l'}{l''}{m}{-m'}{-m''}j_{l''}(kd_{12})Y_{l''m''}
(\hat\bfX_{21})\, .
\end{equation}

Finally we use $\hat\bfX_{21}=-\hat\bfX_{12}$ and {$Y_{l''m''}(-\hat
\bfX)=(-1)^{l''}Y_{l''m''}(\hat \bfX)$}, together with the fact the the $3j$-symbol
restricts $l'+l''-l$ to be even and $m'+m''$ to equal $m$, all of
which yields
\begin{equation}
\label{tr.3.2.1}
\cV^{12}_{{l'm'lm}}(\bfX_{12})=\sqrt{4\pi\lambda\lambda'}\,
i^{l-l'}(-1)^{m}\sum_{l''m''}i^{l''}\sqrt{\lambda''}
\threej{l}{l'}{l''}{0}{0}{0}
\threej{l}{l'}{l''}{m}{-m'}{-m''}j_{l''}(kd_{12})Y_{l''m''}
(\hat\bfX_{12})\, ,
\end{equation}
the analogue of \Eq{forms}. 
 
\subsection{$\cU^{12}_{l'm'lm}(\bfX_{12})$}
\label{app:C2}

We begin by equating the partial wave expansion (with respect to
$\cO_{2}$) of the free Green's function to its plane wave
representation,
\begin{equation}
\label{tr.5}
{\cal G}_{0}(\bfx,\bfx_{2},k)=
ik\sum_{lm}j_{l}(kr)h_{l}^{(1)}(kr_{2})
Y_{lm}^{*}(\hat\bfx)Y_{lm}(\hat\bfx_{2})=
\int\frac{d{\bf q}}{(2\pi)^{3}}\frac{e^{i{\bf q}(\bfx_{2}-\bfx)}}{q^{2}-k^{2}} \, .
\end{equation}
Without loss of generality we take $|\bfx_{2}|>|\bfx|$ with respect to
$\cO_{2}$.  We project out the $(lm)^{\rm th}$ partial wave by
multiplying by $Y_{lm}(\hat\bfx)$ and integrating over $\hat\bfx$,
\begin{equation}
\label{tr.6}
ikj_{l}(kr)h_{l}^{(1)}(kr_2)Y_{lm}({\hat \bfx_2})
=\int d\hat\bfx Y_{lm}(\hat\bfx) \int\frac{d{\bf q}}{(2\pi)^{3}}
\frac{e^{i{\bf q}(\bfx_{2}-\bfx)}}{q^{2}-k^{2}} \, .
\end{equation}
Next we write $e^{i{\bf q}(\bfx_2-\bfx)} = e^{i{\bf
    q}(\bfX_{21}+\bfx_{1}-\bfx)} $ and expand each exponential in
partial waves using \Eq{tr.2}.  The $\hat\bfx$ integral can be
performed using the orthonormality of the $Y_{lm}$; the $\hat{\bf q}$
integral is of the form of \Eq{tr.31}.  The 
{substitution of} $\bfX_{21}=-\bfX_{12}$ proceeds as in the
previous case and we obtain
\begin{eqnarray}
\label{tr.7}
ikj_{l}(kr)h_{l}^{(1)}(kr_2)Y_{lm}({\hat\bfx_2})&=&\frac{8(-1)^{m}}
{\sqrt{4\pi}}\sum_{l'm'l''m''}i^{l+l''-l'}
\sqrt{\lambda\lambda'\lambda''}\threej{l}{l'}{l''}{0}{0}{0}
\threej{l}{l'}{l''}{m}{-m'}{-m''}\nonumber\\
&&\left[\int_{0}^{\infty}dq\frac{q^{2}}{q^{2}-k^{2}}j_{l''}(qd_{12})
j_{l'}(qr_{1})j_{l}(qr)\right]Y_{l''m''}({\hat\bfX_{12}})
Y_{l'm'}({\hat\bfx_{1}})\, .
\end{eqnarray}
This integral can be found in Ref.~\cite{GR},
\begin{equation}
\label{tr.8}
\int_{0}^{\infty}dq\frac{q^{2}}{q^{2}-k^{2}}j_{l''}(q{d_{12}})j_{l'}(q{r_1})j_{l}(qr)
= \frac{ik\pi}{2}j_{l}(kr)j_{l'}(k{r_1})h^{(1)}_{l''}(k{d_{12}}) \, .
\end{equation}
Substituting this {result} into Eq.~\eqref{tr.7}, the form of
$\cU^{12}_{lml'm'}(\bfX_{12})$ quoted in Eq.~(\ref{forms}) follows
straightforwardly.

\end{appendix}

\bibliographystyle{aipprocl}   


\end{document}